\documentclass[12pt,preprint,english,floatfix,showkeys,superscriptaddress,aps,prd]{revtex4}

\usepackage[utf8]{inputenc}
\usepackage[T1]{fontenc}
\usepackage{lmodern}
\usepackage{amsmath}
\usepackage{amssymb}
\usepackage{xcolor}
\usepackage{graphicx}
\usepackage{caption}
\usepackage{subcaption}
\usepackage{float}
\usepackage{esint}
\usepackage{dcolumn}
\usepackage{babel}
\usepackage{csquotes}
\usepackage{color}
\usepackage{slashed}
\usepackage{simplewick}
\usepackage{hyperref}
\hypersetup{
    colorlinks=true,
    citecolor=blue,
    filecolor=green,
    linkcolor=purple,
    urlcolor=red,
}

\usepackage{slashed}

\usepackage{hyperref}
\hypersetup{colorlinks,breaklinks,
			citecolor=[rgb]{0,0.0,1.0},
            urlcolor=[rgb]{0.0,0.0,1.0},
            linkcolor=[rgb]{0,0.5,0.9}}

\begin{document}

\title{Traversable Wormholes Sourced by Dark Matter in Loop Quantum Cosmology}

\author{Marcos V. de S. Silva}
\email{marcosvinicius@fisica.ufc.br}
\affiliation{Departamento de F\'isica, Universidade Federal do Cear\'a, Caixa Postal 6030, Campus do Pici, 60455-760 Fortaleza, Cear\'a, Brazil.}
\author{G. Alencar}
\email{geova@fisica.ufc.br}
\affiliation{Departamento de F\'isica, Universidade Federal do Cear\'a, Caixa Postal 6030, Campus do Pici, 60455-760 Fortaleza, Cear\'a, Brazil.}
\author{R. N. Costa Filho}
\email{rai@fisica.ufc.br}
\affiliation{Departamento de F\'isica, Universidade Federal do Cear\'a, Caixa Postal 6030, Campus do Pici, 60455-760 Fortaleza, Cear\'a, Brazil.}
\author{R. M. P. Neves}
\email{raissa.pimentel@uece.br}
\affiliation{Universidade Estadual do Cear\'a (UECE), Faculdade de Educa\c{c}\~ao, Ci\^encias e Letras de Iguatu, Av. D\'ario Rabelo s/n, Iguatu - CE, 63.500-00 - Brazil.}
\author{Celio R. Muniz}
\email{celio.muniz@uece.br}
\affiliation{Universidade Estadual do Cear\'a (UECE), Faculdade de Educa\c{c}\~ao, Ci\^encias e Letras de Iguatu, Av. D\'ario Rabelo s/n, Iguatu - CE, 63.500-00 - Brazil.}



\date{\today}

\begin{abstract}
In this work, we investigate the existence of wormholes within the framework of Loop Quantum Cosmology, using isotropic dark matter as the source. We analyze three distinct density profiles and solve the modified gravity field equations alongside the stress-energy tensor conservation, applying appropriate boundary conditions to obtain traversable wormhole solutions. Each solution is shown to satisfy the geometric criteria for wormholes, and their regularity is verified by computing the Kretschmann scalar to ensure the absence of singularities under determined conditions. Additionally, we examine the stress-energy tensor to identify scenarios in which energy conditions are violated within this model. The wormhole geometry is further explored through embedding diagrams, and the amount of exotic matter required to sustain these structures is computed using the Volume Integral Quantifier. Finally, we study the shadow produced by our wormhole solution, considering one of the dark matter density profiles, and compare it with observations of the M87 galaxy.

\end{abstract}


\maketitle

\section{Introduction}\label{S:intro}

General Relativity (GR) has successfully described gravitational phenomena for over a century, providing a robust framework for understanding spacetime curvature as influenced by matter and energy. This theory's predictions, from planetary orbits to black hole properties, have been verified extensively, notably in strong-field regimes with the detection of gravitational waves by LIGO and Virgo collaborations \cite{LIGOScientific:2016aoc}. Another fascinating consequence of GR is the possibility of traversable wormholes, hypothetical passages through spacetime that could connect distant points or even different universes \cite{Einstein:1935tc,Morris:1988cz,Hawking:1988ae}. Unlike black holes, which are characterized by an event horizon that prevents anything from escaping, traversable wormholes remain open, allowing for potential passage by particles and light.

However, constructing stable wormhole solutions within GR typically requires exotic matter -- substances that violate energy conditions such as the Null Energy Condition (NEC) \cite{Visser:1995cc,Nandi:2004ku,Churilova:2021tgn,Konoplya:2021hsm}. This need for exotic matter has driven investigations into alternative sources and modifications of GR, particularly in theories that incorporate quantum gravitational effects, such as Loop Quantum Gravity (LQG). Loop Quantum Cosmology (LQC), a simplified model of LQG, introduces corrections to classical GR, especially at high densities, by imposing a critical density \(\rho_c\), beyond which quantum geometric effects become significant \cite{Ashtekar:2008ay}. In this sense, traversable wormholes in scenarios of LQC were built in \cite{Sengupta:2023yof,Muniz:2024jzg}. Such quantum effects hold the potential to reduce or even eliminate the need for exotic matter in sustaining wormholes, allowing other forms of matter, such as dark matter, to support their structure. 

The enigmatic nature of dark matter remains one of physics' greatest puzzles, thought to comprise around five-sixths of the universe's matter \cite{Planck:2018vyg}. Its presence is strongly supported by astrophysical observations, such as galaxy rotation curves that indicate more mass than visible matter alone can explain \cite{Zwicky:1933gu,Rubin:1970zza,Persic:1995ru,Bertone:2016nfn}. Dark matter's mass scale, spanning from cosmic structures down to \(10^{-22} \, \text{eV}\), is still undetermined, making the identification of its particle nature a priority in modern physics \cite{randall2018}. One theory suggests that primordial black holes, originating shortly after the Big Bang, might serve as dark matter. Another possibility involves new particles beyond the Standard Model, like axions and Weakly Interacting Massive Particles (WIMPs), which might accumulate and annihilate in the Sun, emitting neutrinos detectable by observatories like IceCube, though no confirmation has yet been found \cite{Arguelles:2019ouk,Marsh:2024ury}. There are also proposals to detect dark matter in the vicinity of the Sun using satellites orbiting our star \cite{Tsai:2021lly,Souza:2024ltj}. It is worth noting that dark matter as a source of wormholes was investigated in several papers, in Einstein and modified gravity \cite{Xu:2020wfm,Muniz:2022eex,Mustafa:2023kqt,Radhakrishnan:2024rnm,Errehymy:2024lhl,Maurya:2024jos,Hassan:2024xyx}.

In this work, we explore the existence of traversable wormholes sourced by isotropic dark matter within the framework of LQC. We investigate three cold dark matter models -- Navarro-Frenk-White (NFW), Pseudo-Isothermal (PI), and Perfect Fluid (PF) models \cite{Navarro:1995iw,Begeman:1991iy} -- each with distinct density profiles that influence the stability and structure of wormhole solutions. While in the context of General Relativity dark matter does not possess an exotic nature, it may exhibit such characteristics within certain modified gravity frameworks. Then we adopt a linear equation of state, \( p = \omega \rho \), where \(\omega\) can potentially take on negative values. Thus, by employing the LQC-modified Einstein and conservation equations with linear EoS, we derive shape and redshift functions for each dark matter model to assess whether the resulting geometries satisfy the conditions for traversable wormholes, such as the flaring-out condition and asymptotic flatness. Following, we examine the regularity conditions of the corresponding spacetimes by computing the Kretschmann scalar. We further examine energy conditions by analyzing the effective stress-energy tensor derived from the dark matter density profiles, noting that the violation of NEC may vary with the choice of model and parameters, particularly \(\rho_c\) and \(\rho_0\), the LQC critical density and central density of dark matter, respectively.

Through embedding diagrams, we visually represent the wormhole shapes for each model, illustrating how changes in the parameters influence their structure. Furthermore, by calculating the Volume Integral Quantifier (VIQ), we evaluate the amount of exotic matter required to sustain each wormhole, comparing the models between themselves. This work underscores, therefore, the role of quantum corrections from LQC in potentially creating traversable wormholes sustained by dark matter, advancing the study of non-classical spacetimes as viable exotic structures within modified gravity theories. 

The structure of this paper is as follows: In Section II, we introduce the dark matter models to be used and the modified Einstein equations for general traversable wormhole solutions in the context of LQC. In Sections III, IV, and V, we derive these wormhole solutions for the three dark matter models and discuss their geometric properties. Section VI explores the embedding diagrams of the obtained wormhole solutions. In Section VII, we examine the energy conditions and calculate the necessary amount of exotic matter. In Section VIII, we study the shadow formation of our wormhole model based on the NFW dark matter profile, and compare our findings with observations from the Event Horizon Telescope (EHT). Finally, Section IX presents the conclusions and closes the paper.

Throughout this paper, we utilize natural units with $8 \pi G = c =1$ and adopt the metric signature $(-, +, +, +).$

\section{Dark matter models and LQC-Inspired Wormholes }
In this section, we will study general aspects of LQC-Inspired Wormholes sourced by Dark Matter. We will present all the tools necessary to study the explicit cases in the next sections. 

\subsection{Sources and Field Equations}

We will work with the cold dark matter models whose density profiles can be synthesized in the following formula:
\begin{equation}\label{dark matterProfiles}
    \rho(r)=\frac{\rho_0}{\sum_{n=0}^3 a_n (r/R_s)^n},
\end{equation}
where $R_s$ and $\rho_0$ are a distance scale and a density parameter, respectively, which are associated with the dark matter distribution. The coefficients $a_n$ depends on the model under analysis. Thus, we have
\begin{eqnarray}
\text{NFW: }&&a_0=0,\text{ }a_1=1,\text{ }a_2=2,\text{ }a_3=1; \label{nfwDens}\\
\text{PI: }&&a_0=1,\text{ }a_1=0,\text{ }a_2=1,\text{ }a_3=0; \label{piDens}\\
\text{PF: }&&a_0=a_1=a_2=0,\text{ }a_3=1,\label{pfDens}
\end{eqnarray}
where NFW stands for Navarro-Frenk-White, PI is for Pseudo-Isothermal, and PF denotes Perfect Fluid model. A plausible justification for using these dark matter profiles as sources for wormholes in the context of LQC lies in their distinct structural characteristics, which offer a range of gravitational behaviors under some density conditions near the wormhole throat. The NFW profile, with its cuspy core, represents a widely observed density distribution in galaxies and clusters, while the pseudo-isothermal model provides a softened central density, helping to capture alternative galactic dynamics. Meanwhile, the PF dark matter model, often applied in theoretical explorations, features an equation of state that supports pressure, making it particularly useful in analyzing exotic configurations like black holes and wormholes \cite{Li:2012zx,Ashraf:2024bol}. 

Our exploration of traversable wormholes sourced by dark matter is inspired by principles coming from LQC. The solution arises from an effective matter fluid that simulates corrections within the framework of LQC. The modified Friedmann equation of LQC is given by \cite{Swain:2024vnc}
\begin{equation}
H(t)^2 = \frac{\rho}{3}\left(1 - \frac{\rho}{\rho_c}\right),
\end{equation}
for a flat FRW universe, where $\rho_c$ is the critical density that avoids singularities and $H$ is the Hubble parameter. In the late universe, where $H$ becomes approximately constant, the large-scale energy density $\rho$ no longer evolves over time. This justifies the use of a static and local distribution for $\rho$, depending therefore only on position, rather than time. Our model leverages this property to explore these configurations, which are required for wormhole solutions.

Consequently, the effective gravity-matter system obeys the Einstein equations:
\begin{eqnarray}
    G^{\mu}_{\ \nu} \equiv R^{\mu}_{\ \nu} - \frac{1}{2} g^{\mu}_{\ \nu} R =  T^{\mu}_{\ \nu} ,
\end{eqnarray}
where $T_{\mu \nu}$ denotes the effective stress-energy tensor, which is, for an isotropic perfect fluid, given by
\begin{eqnarray}
    T^{\mu}_{\ \nu} = \text{diag}\left(-\rho_{e}, p_{e}, p_{e}, p_{e} \right),
\label{energy_tensor}
\end{eqnarray}
where $\rho_e = - G^{t}_{\ t}$, $p_{e} = G^{r}_{\ r} = G^{\theta}_{\theta} = G^{\phi}_{\phi}$. For a given $\rho$ and $p$, the analytical expressions for the effective energy density and pressure are given by \cite{Sengupta:2023yof,Muniz:2024jzg}:
\begin{eqnarray}
\rho_e(r) &=& \rho\left(1-\frac{\rho}{\rho_c}\right),\label{rhoeff}\\
p_e(r)&=& p-\rho\left(\frac{2p+\rho}{\rho_c}\right),\label{peff}
\end{eqnarray}
For the above expression, we see that even if $p=\omega \rho$  describes normal matter, the effective stress tensor can describe exotic matter. Therefore, it can be a possible source for wormholes. We will discuss this in more detail below. 

Moving forward, we will investigate the wormholes in the context of LQC. We will particularly focus on the energy density given in Eq. \eqref{rhoeff} by the static and spherically symmetric Morris-Thorne wormhole metric as presented by \cite{Morris:1988cz}:
\begin{equation}\label{metric1}
    ds^2=-e^{2\Phi(r)}dt^2+\frac{dr^2}{1-\frac{b(r)}{r}}+r^2d\Omega_2.
\end{equation}
Here, $\Phi(r)$ represents the redshift function, $b(r)$ is the shape function, and $d\Omega_2=d\theta^2+\sin^2\theta d\phi^2$ denotes the spherical line element. Given the metric ansatz of Eq. \eqref{metric1}, the modified Einstein equations take on their simplest form:
\begin{align}
G_{\ t}^{t} = & \frac{b'}{r^{2}}=\rho_e(r),\label{eq:g00}\\
G_{\ r}^{r} = & -\frac{b}{r^{3}}+ 2\frac{\left(r-b\right)\Phi'}{r^{2}}= p_e(r),\label{eq:grr}\\
G_{\theta}^{\theta} = & G_{\phi}^{\phi} = \left(1-\frac{b}{r}\right)\left[\Phi''+(\Phi')^{2}+\frac{\left(b-rb'\right)}{2r(r-b)}\Phi'+\frac{\left(b-rb'\right)}{2r^2(r-b)}+\frac{\Phi'}{r}\right]= p_{e}(r),\label{eq:gthetatheta}
\end{align}
The quantities $\rho(r)$ and $p(r)=\omega \rho(r)$ that entry in the effective densities and pressures are described by Eq. \eqref{dark matterProfiles}, and will be regarded as the sources for the new wormhole solutions investigated here.

Finally, a key feature of the isotropic traversable wormhole solutions to be derived here is that they must satisfy the modified conservation equation:
\begin{equation}\label{TOV} 
\frac{dp_e}{dr} + \frac{d\Phi}{dr}(\rho_e + p_e) = 0, 
\end{equation}
where we assume identical radial and lateral pressures due to the source isotropy. This equation is inherently satisfied, as the Einstein tensor for any general geometry is divergence-free. Equations (\ref{eq:g00}) and (\ref{TOV}),  together with our source, will suffice to find $b(r)$ and $\Phi(r)$. In the following, we find the conditions such that our solutions describe wormholes.

\section{NFW Model}

\subsection{Shape function}
From Eqs. (\ref{dark matterProfiles}), (\ref{nfwDens}), (\ref{rhoeff}), and (\ref{eq:g00}) we find that the corresponding wormhole shape function is given by
\begin{eqnarray}
    b(r)&=&r_0-\frac{R_s^3\rho_0}{\rho_c}\left[\frac{\rho_0}{3\left(1+\frac{r_0}{R_s}\right)^3}+\frac{\rho_c}{\left(1+\frac{r_0}{R_s}\right)}\right]+\frac{ R_s^3\rho_0}{\rho_c}\left[\frac{\rho_0}{3\left(1+\frac{r}{R_s}\right)^3}+\frac{\rho_c}{\left(1+\frac{r}{R_s}\right)}\right]\nonumber\\
    &+& R_s^3\rho_0\log{\left(\frac{1+\frac{r}{R_s}}{1+\frac{r_0}{R_s}}\right)},\label{b_model1}
\end{eqnarray}
where we have taken into account the boundary condition $b(r_0)=r_0$ in order to determine the integration constant.

\begin{figure}
    \centering
    \includegraphics[width=0.49\linewidth]{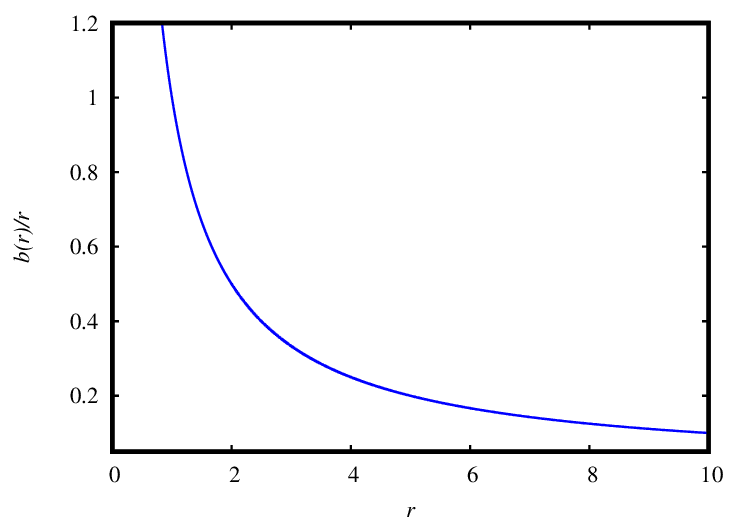}
    \caption{Behavior of $b(r)/r$ considering the model \eqref{b_model1} in terms of the radial coordinate with $r_0=1$, $R_s=2$, $\rho_c=10^{-5}$, and $\rho_0 = 10^{-6}$. Changes in the densities $\rho_c$ and $\rho_0$ do not significantly alter the shape of the graph.}
    \label{fig:br_model1}
\end{figure}

\begin{figure}
    \centering
    \includegraphics[width=0.49\linewidth]{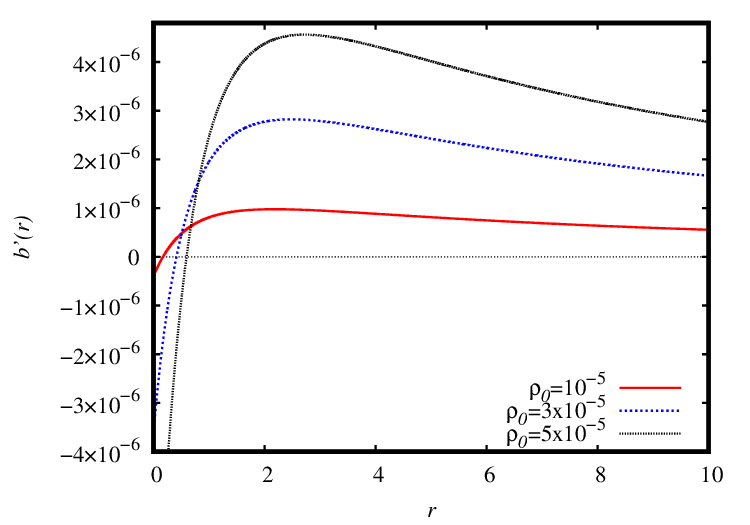}
    \caption{Behavior of $b'(r)$ considering the model \eqref{b_model1} in terms of the radial coordinate with $r_0=1$, $R_s=2$, and $\rho_c=10^{-5}$, for different values of $\rho_0$. Changes in the density $\rho_c$ do not significantly alter the shape of the graph.}
    \label{fig:bline_model1}
\end{figure}

\begin{figure}
    \centering
    \includegraphics[width=0.49\linewidth]{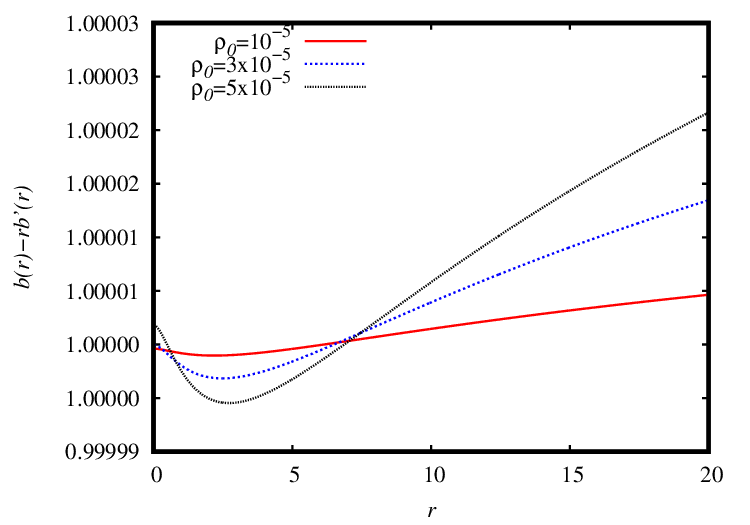}
    \caption{Behavior of $b(r)-rb'(r)$ considering the model \eqref{b_model1} in terms of the radial coordinate with $r_0=1$, $R_s=2$, and $\rho_c=10^{-5}$, for different values of $\rho_0$. Changes in the density $\rho_c$ do not significantly alter the shape of the graph.}
    \label{fig:bflaring_model1}
\end{figure}

The condition $b(r_0)=r_0$ was used to integrate the field equation and obtain the form of $b(r)$. However, we still need to verify whether the other conditions mentioned in the before are also satisfied. In Fig. \ref{fig:br_model1}, we observe the behavior of $b(r)/r$ and notice that, for $r>r_0$, the ratio $b(r)/r$ is always less than one, thus satisfying the condition \textit{ii} that we established earlier. In Fig. \ref{fig:bline_model1}, we analyze the behavior of $b'(r)$ and observe that it is always less than one, ensuring that the condition $b'(r_0) < 1$ is easily satisfied. In Fig. \ref{fig:bflaring_model1}, we examine the behavior of the relation $b(r) - r b'(r)$  and observe that it is always positive, ensuring that the flaring condition is always satisfied. Thus, we see that the wormhole model derived from LQC, when considering dark matter described by the NFW profile, satisfies the established conditions.

\subsection{Redshift function}
\begin{figure}
    \centering
    \includegraphics[width=0.5\linewidth]{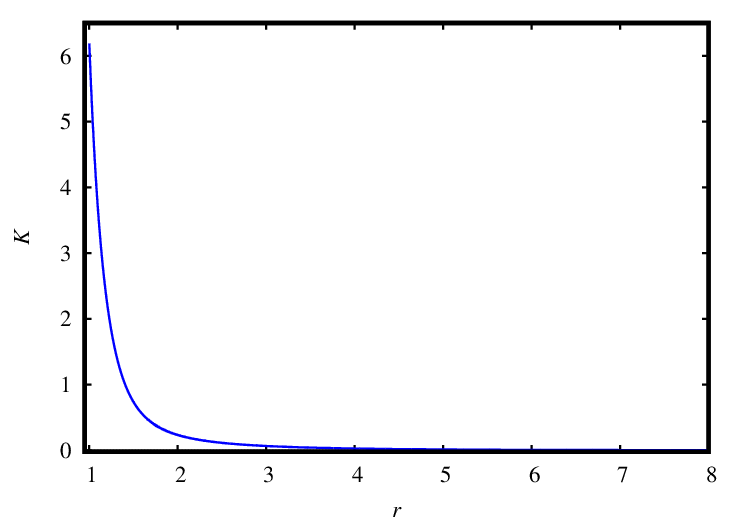}
    \caption{Behavior of the Kretschmann scalar for model \eqref{b_model1}, as a functions of the radial coordinate, with $\omega=1$, $r_0 = 1$, $R_s = 2$, $\rho_c = 10^{-5}$, and $\rho_0 = 10^{-6}$.}
    \label{fig:K-Model1}
\end{figure}

From Eq. (\ref{dark matterProfiles}),  (\ref{nfwDens}), (\ref{TOV}), and the state equation $p(r)=\omega \rho(r)$, we find
\begin{equation}\label{NFWredshift}
    e^{2\Phi(r)}=\left(\frac{r}{r_0}\right)^{\frac{2(1+2\omega)}{1+\omega}}\left(\frac{r+R_s}{r_0+R_s}\right)^{\frac{4(1+2\omega)}{1+\omega}}\left[\frac{r_0(r_0+R_s)^2\rho_c-2R_s^3\rho_0}{r(r+R_s)^2\rho_c-2R_s^3\rho_0}\right]^2,
\end{equation}
where we have chosen the integration constant such that, at the throat, $e^{2\Phi(r_0)}=1$. The time coefficient of this metric exhibits an undesirable asymptotic behavior, except when $\omega\to 0$, thus requiring, in general, the imposition of junction conditions. However, the curvature scalars clearly indicate that asymptotic flatness is achieved. This can be seen in Fig. \ref{fig:K-Model1}, where we graphically observe the behavior of the Kretschmann scalar and see that it does not show divergences as it approaches the throat radius, decreasing to zero as the radial coordinate increases. Respecting the regularity condition imposed on the energy density of the dark matter and considering that $\rho_c$ must have small values to deviate from the general relativity results, parameter changes do not significantly alter the behavior of the Kretschmann scalar; therefore, we have not included additional curves in Fig. \ref{fig:K-Model1}. as we will demonstrate. This property holds consistently across all dark-matter models. Despite the Kretschmann being quite lengthy, it will give us simple conditions for the absence of singularities. 

First of all, we note that our throat is at $r=r_0$, therefore, factors of $1/r^n$ are not singular. From the conditions satisfied by the shape function,  we have that 
\begin{equation}\label{condr}
b'(r)<\frac{b(r)}{r} < 1,    
\end{equation}
and $b(r)$ will not contribute to singularities. We find, with this, that the singularities can appear solely from the factors of $\Phi ''(r)$ and $ \Phi '(r)$ present in the Kretschmann scalar. Therefore, in order to avoid singularities, we must impose that
\begin{equation}\label{singcond}
    \Phi ''(r), \Phi '(r)\neq \infty \quad \forall \, r >r_0 .
\end{equation} 
\begin{align}
  \Phi '(r)=&\frac{
    (3r + R_s) \left( \rho_c r^3 \omega + 2 \rho_c r^2 R_s \omega + \rho_c r R_s^2 \omega - 2 R_s^3 (2 \rho_0 \omega + \rho_0) \right)
}{
    r (\omega + 1) (r + R_s) \left( \rho_c r (r + R_s)^2 - 2 \rho_0 R_s^3  \right)
},\\
  \Phi ''(r)=&\frac{
    \omega \rho_c^2 (r + R_s)^2 (3r + R_s)^2
}{
    (1 + \omega) 
    \left( 
        \rho_c r (r + R_s)^2 - 2 \rho_0 R_s^3 
    \right)^2
}-\frac{
    \omega (6 \rho_c r + 4 \rho_c R_s)
}{
    (1 + \omega) \left( \rho_c r (r + R_s)^2 - 2 \rho_0 R_s^3 \right)
}
\\
&+\frac{
    \rho_c^2 (r + R_s)^2 (3r + R_s)^2
}{
    (1 + \omega) 
    \left( 
        \rho_c r (r + R_s)^2 - 2 \rho_0 R_s^3 
    \right)^2
}+\frac{
    6 \rho_c r + 4 \rho_c R_s
}{
    (1 + \omega) \left( \rho_c r (r + R_s)^2 - 2 \rho_0 R_s^3 \right)
}
\\
& +\frac{4 \omega}{(1 + \omega) (r + R_s)^2}
+ \frac{2}{(1 + \omega) (r + R_s)^2}
+ \frac{2 \omega + 1}{(1 + \omega) r^2}.
\end{align}
From the above expressions, we see that 
\begin{equation}
   \lim_{r \to \infty} \Phi ''(r) = \lim_{r \to \infty}\Phi '(r) = 0,
\end{equation}
and this is enough to guarantee asymptotic flatness. We also see that our spacetime will only be free of singularities for $\rho_0<\rho_c \, r_0 \, (r_0 + R_s)^2/2 \, R_s^3$. Thus, we verify that if the dark matter density is very high, it may end up generating singularities in our spacetime.

\section{PI Model}

\subsection{Shape function}

\begin{figure}
    \centering
    \includegraphics[width=0.49\linewidth]{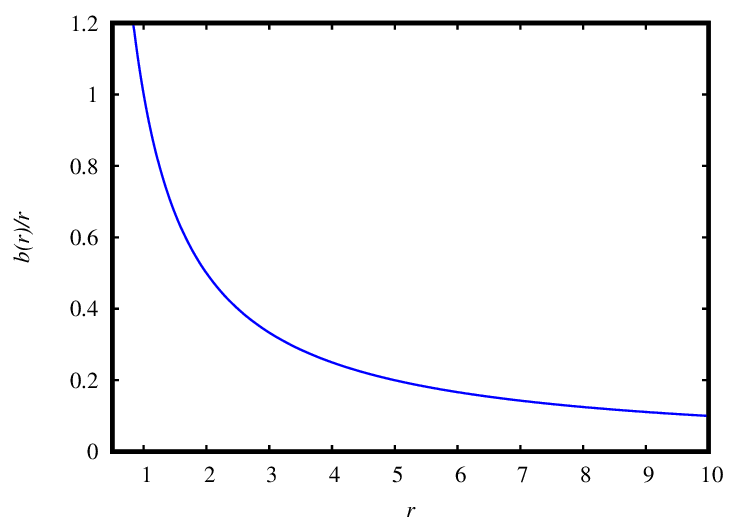}
    \caption{Behavior of $b(r)/r$ considering the model \eqref{b_model2} in terms of the radial coordinate with $r_0=1$, $R_s=2$, $\rho_c=5\times10^{-5}$, and $\rho_0 = 10^{-6}$. Changes in the densities $\rho_c$ and $\rho_0$ do not significantly alter the shape of the graph.}
    \label{fig:br_model2}
\end{figure}

\begin{figure}
    \centering
    \includegraphics[width=0.49\linewidth]{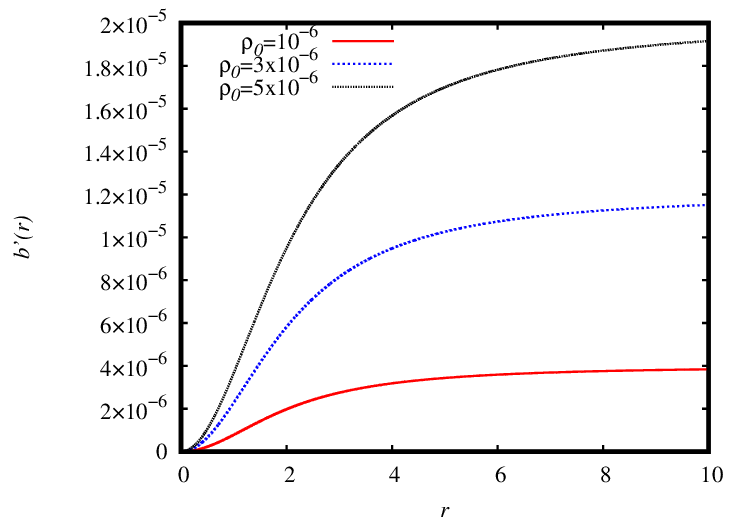}
    \caption{Behavior of $b'(r)$ considering the model \eqref{b_model2} in terms of the radial coordinate with $r_0=1$, $R_s=2$, and $\rho_c=5\times10^{-5}$, for different values of $\rho_0$. Changes in the density $\rho_c$ do not significantly alter the shape of the graph.}
    \label{fig:bline_model2}
\end{figure}

\begin{figure}
    \centering
    \includegraphics[width=0.49\linewidth]{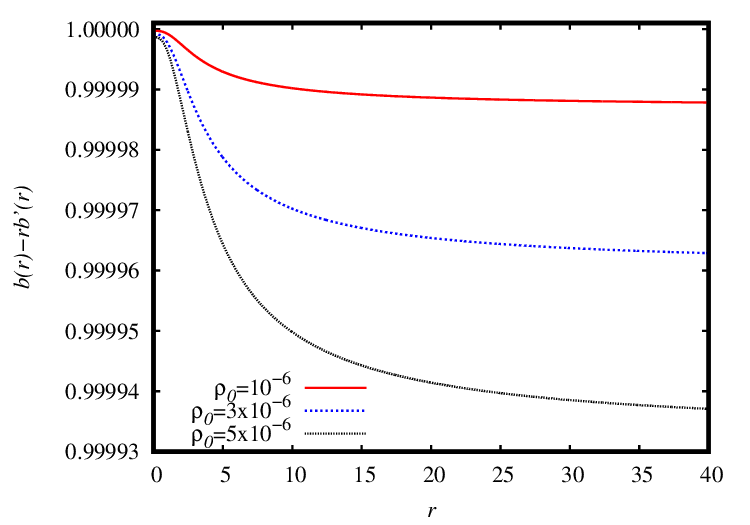}
    \caption{Behavior of $b(r)-rb'(r)$ considering the model \eqref{b_model2} in terms of the radial coordinate with $r_0=1$, $R_s=2$, and $\rho_c=5\times10^{-5}$, for different values of $\rho_0$. Changes in the density $\rho_c$ do not significantly alter the shape of the graph.}
    \label{fig:bflaring_model2}
\end{figure}

From Eqs. (\ref{dark matterProfiles}), (\ref{piDens}), (\ref{rhoeff}), and (\ref{eq:g00}) we find that the corresponding wormhole shape function is given by
\begin{eqnarray}
    b(r)&=&r_0- \rho_0 r_0 R_s^2-\frac{  \rho_0^2 r_0 R_s^4}{2\rho_c } \left(r_0^2+R_s^2\right)+\rho_0 R_s^3 \tan ^{-1}\left(\frac{r_0}{R_s}\right)+\frac{ \rho_0^2 R_s^3 \tan ^{-1}\left(\frac{r_0}{R_s}\right)}{2\rho_c }\nonumber\\
    &+&  \rho_0 r R_s^2+\frac{  \rho_0^2 r R_s^4}{2\rho_c \left(r^2+R_s^2\right)}-\frac{ \rho_0^2 R_s^3 \tan ^{-1}\left(\frac{r}{R_s}\right)}{2\rho_c }-  \rho_0 R_s^3 \tan ^{-1}\left(\frac{r}{R_s}\right).\label{b_model2}
\end{eqnarray}
Let us now verify whether the model satisfies the conditions imposed previously. The condition $b(r_0)=r_0$ is automatically satisfied since this condition was imposed during the integration process to obtain $b(r)$. The second condition, $b(r)/r<1$ for $r > r_0$, is always satisfied, as we see from Fig. \ref{fig:br_model2} that for model \eqref{b_model2} the function $b(r)/r$ is always less than one for $r > r_0$. Through Fig. \ref{fig:bline_model2}, we see that $b'(r)$ is always less than one, even taking negative values. This shows that the third condition, $b'(r_0) < 1$, is satisfied. Through Fig. \ref{fig:bflaring_model2}, we see that $b(r) - b'(r)r > 0$ for model \eqref{b_model2}, so that the fourth condition is always satisfied. Thus, depending on the chosen parameter values, model \eqref{b_model2} meets all the conditions for a wormhole.

\subsection{Redshift function}
\begin{figure}
    \centering
    \includegraphics[width=0.5\linewidth]{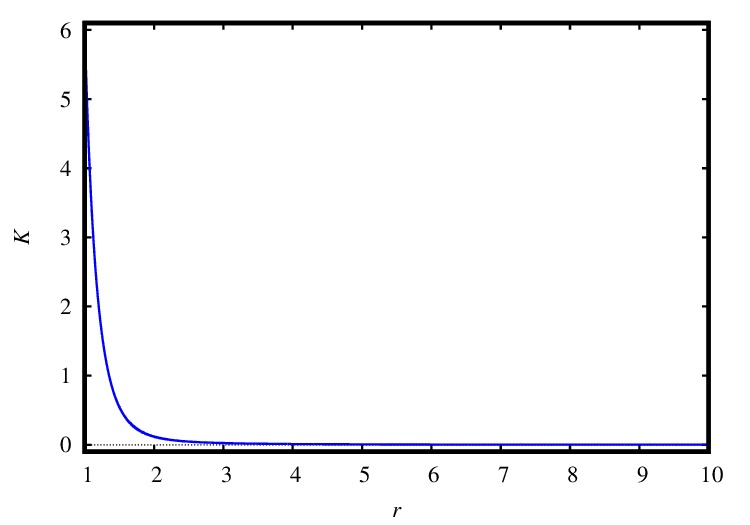}
    \caption{Behavior of the Kretschmann scalar for model \eqref{b_model2}, as a functions of the radial coordinate, with $\omega=1$, $r_0 = 1$, $R_s = 2$, $\rho_c = 5\times 10^{-6}$, and $\rho_0 = 10^{-6}$.}
    \label{fig:K-Model2}
\end{figure}

From Eq. (\ref{dark matterProfiles}),  (\ref{piDens}), (\ref{TOV}), and state equation $p(r)=\omega \rho(r)$, we find
\begin{equation}
    e^{2\Phi(r)}=\left(\frac{r^2+Rs^2}{r_0^2+Rs^2}\right)^{\frac{2(1+2\omega)}{1+\omega}}\left[\frac{(r_0^2+Rs^2)\rho_c-2Rs^2\rho_0}{(r^2+Rs^2)\rho_c-2Rs^2\rho_0}\right]^2.
\end{equation}
Once we have the shape function and the redshift function, we are able to calculate the Kretschmann scalar for the second model. However, as in the previous case, the analytical expression is not clear.
\begin{align}
     \Phi '(r)=&  
    \frac{2r\rho_c}{2 \rho_0 R_s^2 - \rho_c \left( r^2 + R_s^2 \right)} 
    + \frac{2r(2 \omega + 1)}{(\omega + 1) \left( r^2 + R_s^2 \right)},
\\
 \Phi ''(r)=&\frac{4 \rho_c^2 r^2}{\left( \rho_c r^2 + R_s^2 (\rho_c - 2 \rho_0) \right)^2}
- \frac{2 \rho_c}{\rho_c r^2 + R_s^2 (\rho_c - 2 \rho_0)}
- \frac{2 (2 \omega + 1) \left( r^2 - R_s^2 \right)}{(\omega + 1) \left( r^2 + R_s^2 \right)^2}.
\end{align}
An information we can extract from the analytical expression is that there are no singularities if $\rho_0 < \rho_c / 2$. This limits the possible values of the dark matter energy density. In Fig. \ref{fig:K-Model2}, we graphically observe the behavior of the Kretschmann scalar and see that it does not show divergences as it approaches the throat radius, decreasing to zero as the radial coordinate increases. As in the case before, parameter changes do not significantly alter the behavior of the Kretschmann scalar; therefore, we have not included additional curves in Fig. \ref{fig:K-Model2}.

\section{PF Model}\label{S:PF}

\subsection{Shape function}

\begin{figure}
    \centering
    \includegraphics[width=0.49\linewidth]{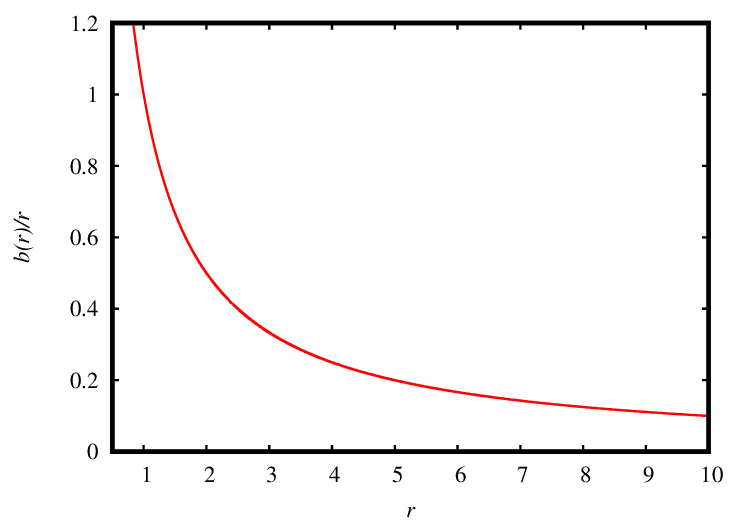}
    \caption{Behavior of $b(r)/r$ considering the model \eqref{b_model3} in terms of the radial coordinate with $r_0=1$, $R_s=2$, $\rho_c=5\times10^{-5}$, and $\rho_0 = 10^{-7}$. Changes in the densities $\rho_c$ and $\rho_0$ do not significantly alter the shape of the graph.}
    \label{fig:br_model3}
\end{figure}

\begin{figure}
    \centering
    \includegraphics[width=0.49\linewidth]{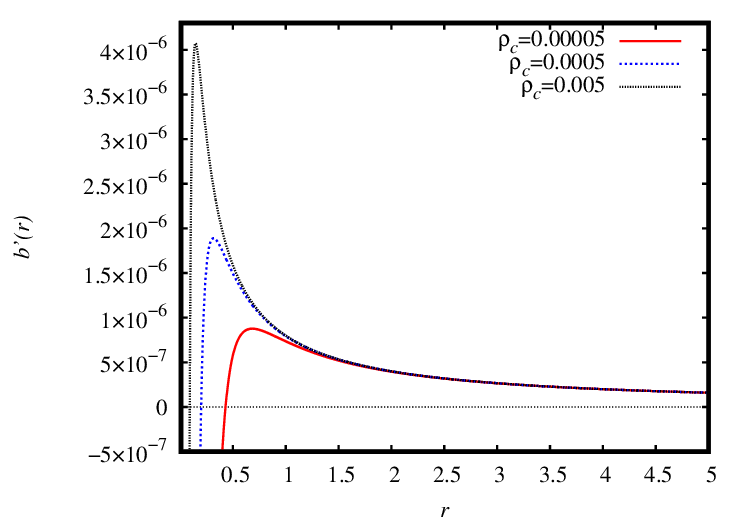}
    \includegraphics[width=0.49\linewidth]{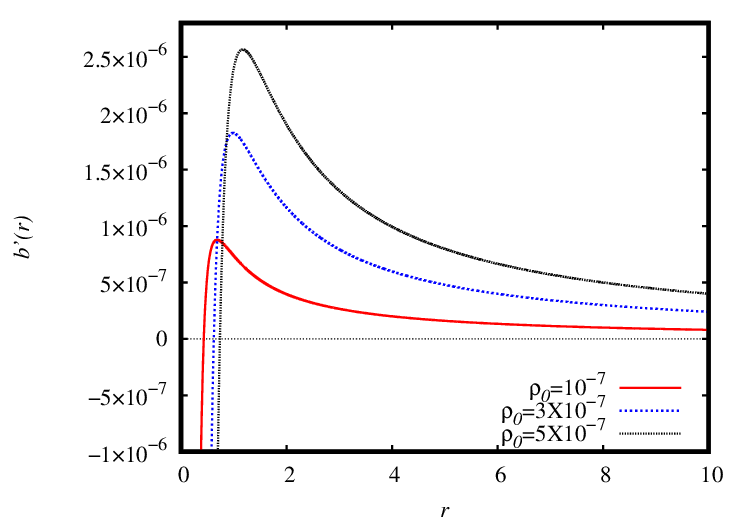}
    \caption{Behavior of $b'(r)$ considering the model \eqref{b_model3} in terms of the radial coordinate with $r_0=1$ and $R_s=2$. In the left panel, we fix $\rho_0=10^{-7}$ and vary $\rho_c$. In the right panel, we fix $\rho_c=5\times 10^{-5}$ and vary $\rho_0$.}
    \label{fig:bline_model3}
\end{figure}

\begin{figure}
    \centering
    \includegraphics[width=0.49\linewidth]{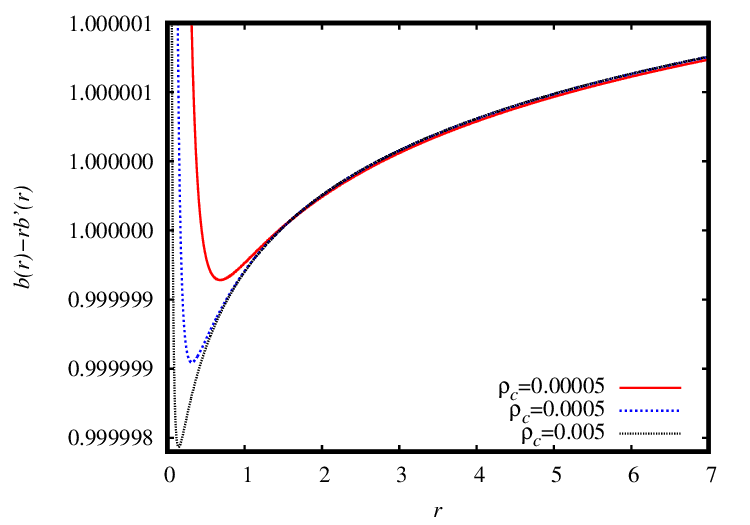}
    \includegraphics[width=0.49\linewidth]{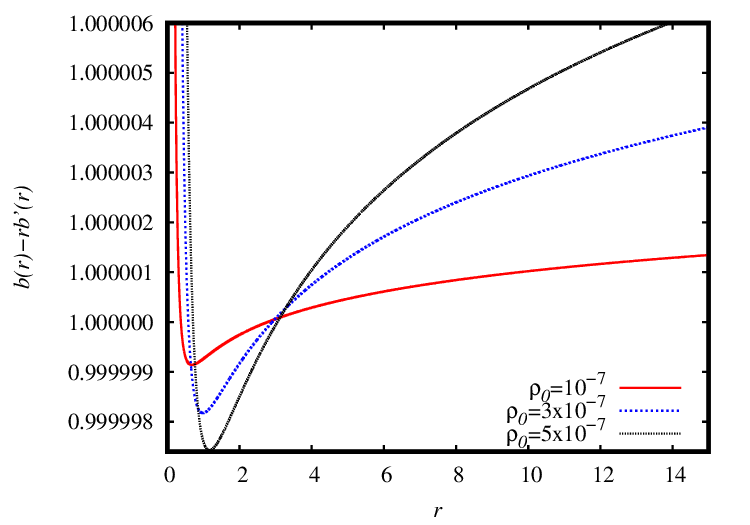}
    \caption{Behavior of $b(r)-rb'(r)$ considering the model \eqref{b_model3} in terms of the radial coordinate with $r_0=1$ and $R_s=2$. In the left panel, we fix $\rho_0=10^{-7}$ and vary $\rho_c$. In the right panel, we fix $\rho_c=5\times 10^{-5}$ and vary $\rho_0$.}
    \label{fig:bflaring_model3}
\end{figure}

From Eqs. (\ref{dark matterProfiles}), (\ref{pfDens}), (\ref{rhoeff}), and (\ref{eq:g00}), we find that the corresponding wormhole shape function is given by
\begin{eqnarray}
    b(r)=r_0-\frac{ \rho_0^2 R_s^6}{3 \rho_c r_0^3}+\frac{ \rho_0^2 R_s^6}{3 \rho_c r^3}+ \rho_0 R_s^3 \log{\left(\frac{r}{r_0}\right)}.\label{b_model3}
\end{eqnarray}
As in the previous cases, the condition $b(r_0) = r_0$ is identically satisfied since it was used in the integration to obtain $b(r)$. From Fig. \ref{fig:br_model3}, we see that $b(r)/r < 1$ for $r > r_0$, which satisfies the second condition for wormholes. In Fig. \ref{fig:bline_model3}, we observe that $b' < 1$ at all points with different parameter values, which ensures that $b'(r_0) < 1$. Finally, in Fig. \ref{fig:bflaring_model3}, we analyze the behavior of $b(r) - b'(r)r$ and verify that it is always positive for the chosen parameters. Thus, the wormhole that arises due to the dark matter profile \ref{pfDens} in LQC satisfies all the conditions imposed for wormholes.

\subsection{Redshift function}
From Eq. (\ref{dark matterProfiles}),  (\ref{pfDens}), (\ref{TOV}), and state equation $p(r)=\omega \rho(r)$, we find
\begin{equation}
    e^{2\Phi(r)}=\left(\frac{r}{r_0}\right)^{\frac{6(1+2\omega)}{1+\omega}}\left(\frac{r_0^3\rho_c-2Rs^3\rho_0}{r^3\rho_c-2Rs^3\rho_0}\right)^2,
\end{equation}
As in previous cases, we now have the necessary functions to calculate the Kretschmann scalar. Although the analytical expression is simpler than in previous cases, it is still quite confusing and unclear.
\begin{align}
   \Phi '(r)=&   \frac{3 \rho_c r^2}{2 \rho_0 R_s^3 - \rho_c r^3}
+ \frac{6 \omega + 3}{r \omega + r},
\\
 \Phi ''(r)=& -\frac{
    3 \left( \rho_c^2 r^6 \omega - 4 \rho_0 \rho_c r^3 R_s^3 (3 \omega + 2) + 4 \rho_0^2 R_s^6 (2 \omega + 1) \right)
}{
    (\omega + 1) \left( \rho_c r^4 - 2 \rho_0 r R_s^3 \right)^2
}.
\end{align}

\begin{figure}
    \centering
    \includegraphics[width=0.5\linewidth]{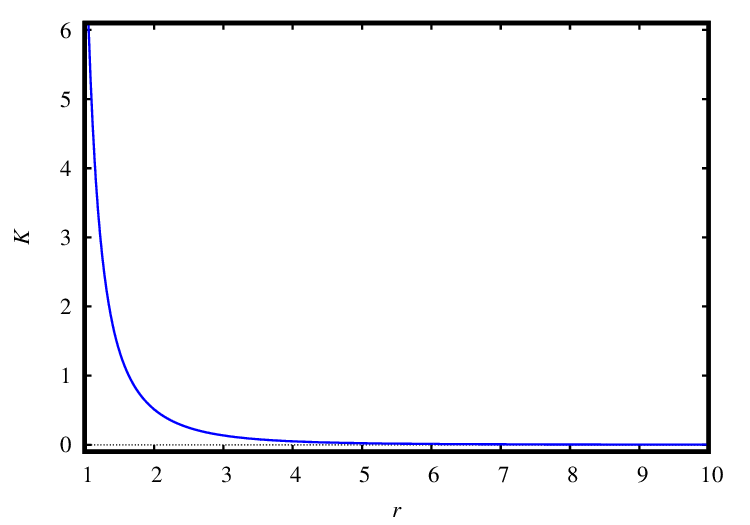}
    \caption{Behavior of the Kretschmann scalar for model \eqref{b_model3}, as a functions of the radial coordinate, with $\omega=1$, $r_0 = 1$, $R_s = 2$, $\rho_c = 5\times 10^{-5}$, and $\rho_0 = 10^{-7}$.}
    \label{fig:K-Model3}
\end{figure}

Nevertheless, we can verify that, for our spacetime to be free of singularities, we need $\rho_0 < r_0^3 \, \rho_c/2 \, R_s^3$. In Fig. \ref{fig:K-Model3}, we graphically observe the behavior of the Kretschmann scalar and see that it does not show divergences as it approaches the throat radius, decreasing to zero as the radial coordinate increases. As in the cases before, parameter changes do not significantly alter the behavior of the Kretschmann scalar; therefore, we have not included additional curves in Fig. \ref{fig:K-Model3}.

\section{Embedding diagrams}\label{S:embedding}
In order to better analyze the shape of the wormholes we found, we will study the embedding diagrams. These type of diagrams help us to understand the curvature of the spacetime around compact objects, such as the wormholes.

To construct these diagrams, we will consider the line element that describes our wormhole models for the case $t = \text{constant}$ and considering the equatorial plane, $\theta = \pi/2$. With these simplifications, the line element is written as:
\begin{equation}\label{metric2d}
    ds^2=\frac{dr^2}{1-\frac{b(r)}{r}}+r^2d\phi^2.
\end{equation}
Now, we will embed this spacetime into another spacetime, which is three-dimensional with cylindrical symmetry, assuming $\theta = \pi/2$. This three-dimensional spacetime is described by the line element:
\begin{equation}
    ds^2_{cyl}=d\rho^2+\rho^2d\phi^2+dz^2.
\end{equation}
Comparing the two line elements, we can identify:
\begin{equation}
    \rho=r, \quad \mbox{and} \quad \frac{dz}{dr}=\pm \left(\frac{r}{b}-1\right)^{-1/2}.\label{Zrgen}
\end{equation}
Now, we just need to integrate equation \eqref{Zrgen} to obtain the shape of these wormholes to each model.

\begin{figure}
    \centering
    \includegraphics[width=0.49\linewidth]{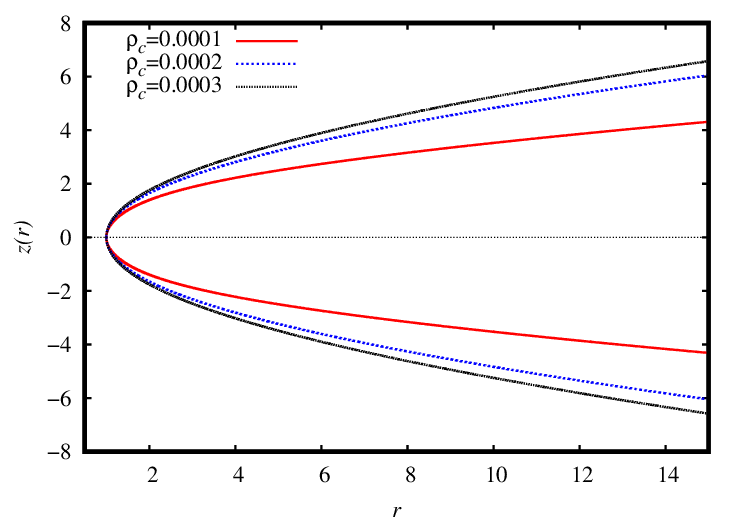}
    \includegraphics[width=0.49\linewidth]{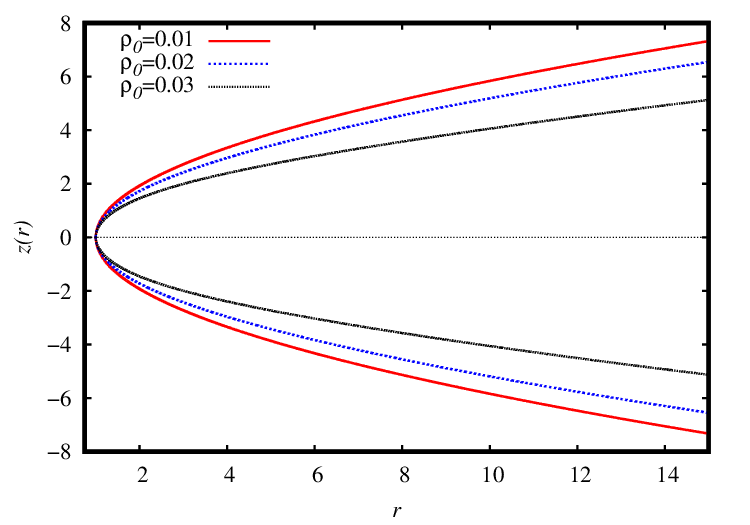}
    \caption{Behavior of $z(r)$ considering the model \eqref{b_model1} in terms of the radial coordinate with $r_0=1$ and $R_s=2$. In the left panel, we fix $\rho_0=0.01$ and vary $\rho_c$. In the right panel, we fix $\rho_c=0.001$ and vary $\rho_0$.}
    \label{fig:Zr_model1}
\end{figure}

\begin{figure}
    \centering
    \includegraphics[width=0.49\linewidth]{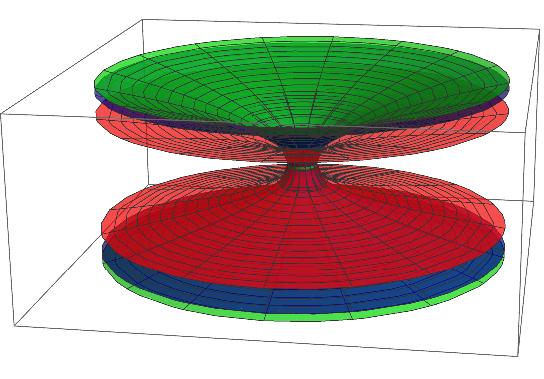}
    \includegraphics[width=0.49\linewidth]{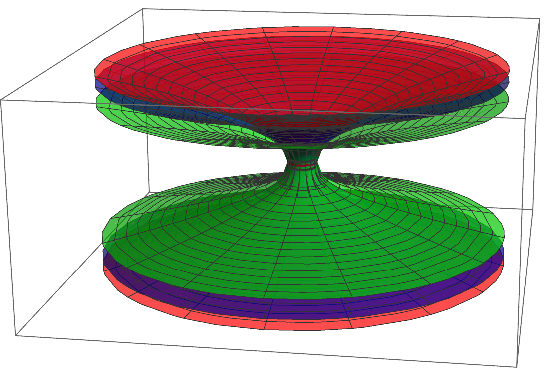}
    \caption{Embedding diagrams considering the model \eqref{b_model1} in terms of the radial coordinate with $r_0=1$ and $R_s=2$. In the left panel, we fix $\rho_0=0.01$ and consider $\rho_c=0.0001$ (green), $\rho_c=0.0002$ (blue), and $\rho_c=0.0003$ (red). In the right panel, we fix $\rho_c=0.001$ and consider $\rho_0=0.01$ (red), $\rho_0=0.02$ (blue), and $\rho_0=0.03$ (green).}
    \label{fig:Zrev_model1}
\end{figure}

In Figs. \ref{fig:Zr_model1} and \ref{fig:Zrev_model1}, we show the behavior of the embedding diagram profile and the three-dimensional version of the diagrams for model \eqref{b_model1}. Depending on the chosen parameter values, the wormhole flattens more quickly. The higher the value of $\rho_0$ or the lower the value of $\rho_c$, the faster the wormhole flattens.

\begin{figure}
    \centering
    \includegraphics[width=0.49\linewidth]{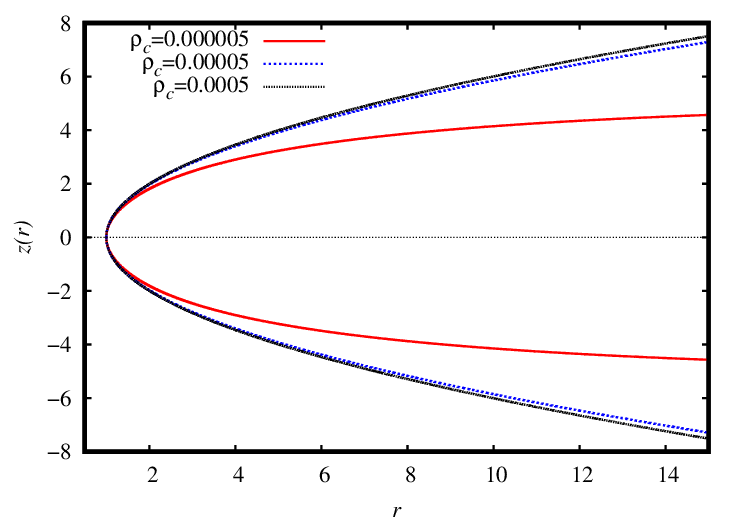}
    \includegraphics[width=0.49\linewidth]{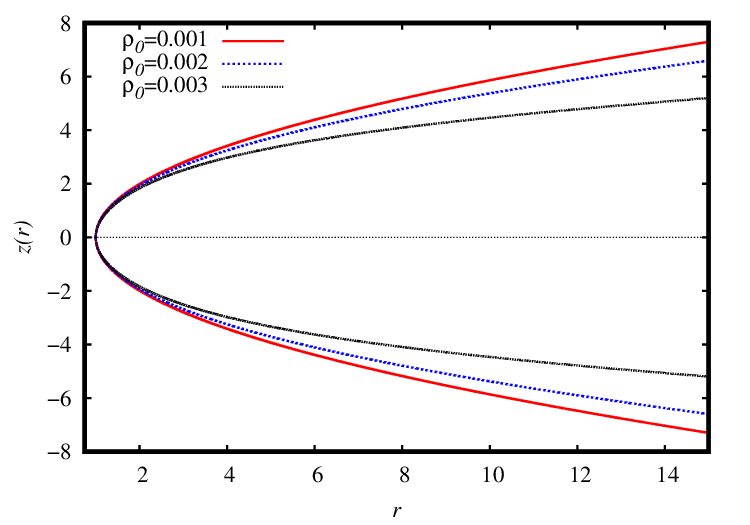}
    \caption{Behavior of $z(r)$ considering the model \eqref{b_model2} in terms of the radial coordinate with $r_0=1$ and $R_s=2$. In the left panel, we fix $\rho_0=0.001$ and vary $\rho_c$. In the right panel, we fix $\rho_c=0.00005$ and vary $\rho_0$.}
    \label{fig:Zr_model2}
\end{figure}

\begin{figure}
    \centering
    \includegraphics[width=0.49\linewidth]{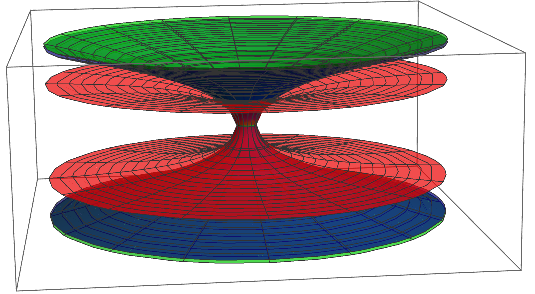}
    \includegraphics[width=0.49\linewidth]{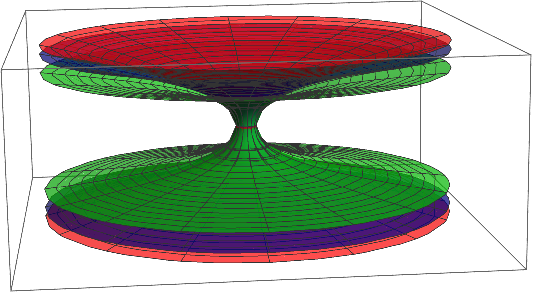}
    \caption{Embedding diagrams considering the model \eqref{b_model2} in terms of the radial coordinate with $r_0=1$ and $R_s=2$. In the left panel, we fix $\rho_0=0.001$ and consider $\rho_c=0.000005$ (green), $\rho_c=0.00005$ (blue), and $\rho_c=0.0005$ (red). In the right panel, we fix $\rho_c=0.00005$ and consider $\rho_0=0.001$ (red), $\rho_0=0.002$ (blue), and $\rho_0=0.003$ (green).}
    \label{fig:Zrev_model2}
\end{figure}

In Figs. \ref{fig:Zr_model2} and \ref{fig:Zrev_model2}, we show the behavior of the embedding diagram profile and the three-dimensional version of the diagrams for model \eqref{b_model2}. The behavior is similar to the first model, where the higher the value of $\rho_0$ or the lower the value of $\rho_c$, the faster the solution flattens. It is interesting to note that for larger values of $\rho_c$, the shape of the wormhole does not change much when $\rho_c$ is further increased.

\begin{figure}
    \centering
    \includegraphics[width=0.49\linewidth]{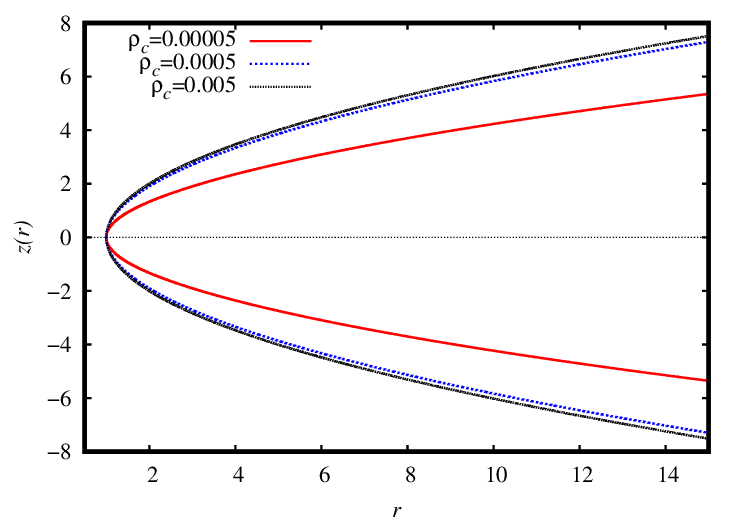}
    \includegraphics[width=0.49\linewidth]{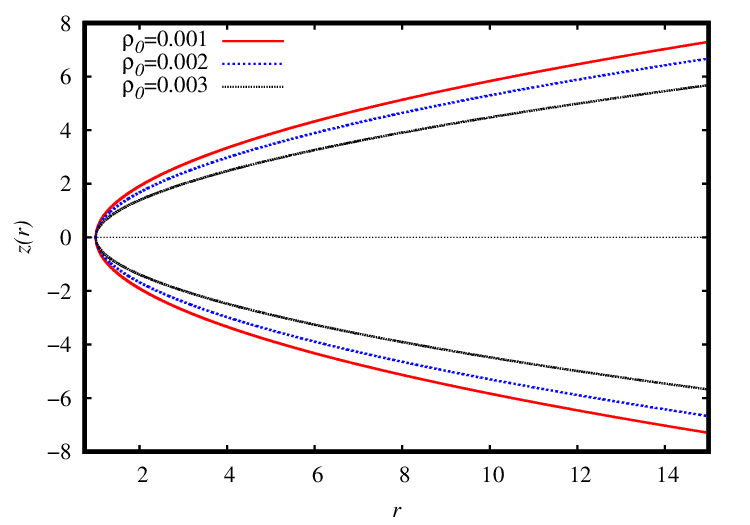}
    \caption{Behavior of $z(r)$ considering the model \eqref{b_model3} in terms of the radial coordinate with $r_0=1$ and $R_s=2$. In the left panel, we fix $\rho_0=0.001$ and vary $\rho_c$. In the right panel, we fix $\rho_c=0.0005$ and vary $\rho_0$.}
    \label{fig:Zr_model3}
\end{figure}

\begin{figure}
    \centering
    \includegraphics[width=0.49\linewidth]{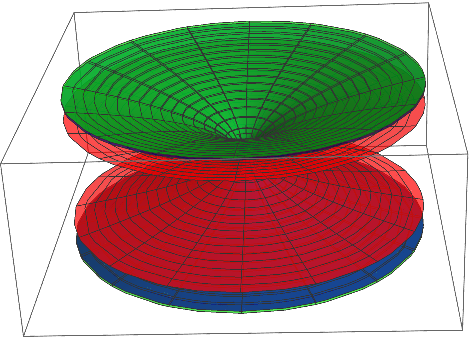}
    \includegraphics[width=0.49\linewidth]{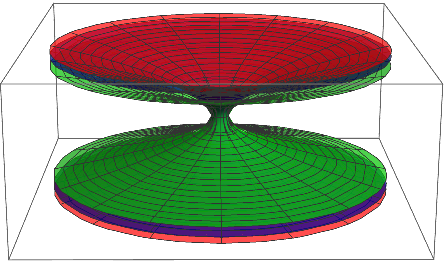}
    \caption{Embedding diagrams considering the model \eqref{b_model3} in terms of the radial coordinate with $r_0=1$ and $R_s=2$. In the left panel, we fix $\rho_0=0.001$ and consider $\rho_c=0.00005$ (green), $\rho_c=0.0005$ (blue), and $\rho_c=0.005$ (red). In the right panel, we fix $\rho_c=0.0005$ and consider $\rho_0=0.001$ (red), $\rho_0=0.002$ (blue), and $\rho_0=0.003$ (green).}
    \label{fig:Zrev_model3}
\end{figure}

Finally, \ref{fig:Zr_model3} and \ref{fig:Zrev_model3}, we show the behavior of the embedding diagram profile and the three-dimensional version of the diagrams for model \eqref{b_model3}. The behavior is similar to the first model, where the higher the value of $\rho_0$ or the lower the value of $\rho_c$, the faster the solution flattens. Thus, even though the shape function of each wormhole is completely different, graphically the behavior of the wormhole shape for each dark matter profile is quite similar.

\section{Energy Conditions and amount of exotic matter}\label{S:energy_conditions}
The energy conditions are constraints imposed on the components of a given stress-energy tensor to ensure its physical viability. In the case of wormholes, the violation of certain energy conditions is essential to keep the wormhole's throat traversable.

To our case, let's consider the components of the stress-energy tensor given by \eqref{energy_tensor}. The energy conditions are given by the inequalities:
\begin{eqnarray}
&&NEC=WEC_{1}=SEC_{1} 
\Longleftrightarrow \rho_e+p_{e}\geq 0,\label{Econd1} \\
&&SEC_2 \Longleftrightarrow\rho_e+3p_e\geq 0,\label{Econd2}\\
&&DEC_{1} \Longleftrightarrow \rho_e-|p_{e}|\geq 0 \Longleftrightarrow 
(\rho_e+p_{e}\geq 0) \hbox{ and } (\rho_e-p_{e}\geq 0),\label{Econd3}\\
&&DEC_2=WEC_2 \Longleftrightarrow\rho_e\geq 0,\label{Econd4}
\end{eqnarray}
where the energy conditions are the null (NEC), weak (WEC), strong (SEC), and dominant (DEC). We see that the NEC is embedded within the other conditions. Thus, if the NEC is violated, the other conditions will also be violated. However, the reverse is not necessarily true, as the other conditions also impose additional inequalities that do not appear in the NEC. Now let's analyze the conditions for each dark matter model.

In this section, we will also analyze the quantity of exotic matter necessary to maintain the wormhole by calculating the Volume Integral Quantifier (VIQ), defined by \cite{Nandi:2004ku}
\begin{equation}\label{viq}
\mathcal{I}_v=\oint 4\pi r^2 (\rho_e+p_e)dr = 2\int_{r_0}^r 4\pi x^2 (\rho_e+p_e)dx,
\end{equation}
where $r\to \infty$. We will perform this calculation for the three dark matter models considered up to this point.

\subsection{NFW model}
Considering the model \eqref{nfwDens}, we can calculate the components of the stress-energy tensor, and we obtain that their combinations are given by:
\begin{eqnarray}
&&\rho_e+p_{e}=\frac{\rho_0 R_s^3 (\omega +1) \left(r\rho_c(r+R_s)^2-2 \rho_0 R_s^3\right)}{\rho_c r^2 (r+R_s)^4},\label{Econd1_model1} \\
&&\rho_e+3p_e=\frac{\rho_0 R_s^3 \left(r\rho_c(r+ R_s )^2(3 \omega +1)-2 \rho_0 R_s^3 (3 \omega +2)\right)}{\rho_c r^2
   (r+R_s)^4},\label{Econd2_model1}\\
&& \rho_e-p_{e}=\frac{\rho_0 R_s^3 \left(r\rho_c(r+ R_s)^2(1-\omega)+2 \rho_0 R_s^3 \omega \right)}{\rho_c r^2
   (r+R_s)^4},\label{Econd3_model1}\\
&&\rho_e=\frac{\rho_0 R_s^3 \left(\rho_c r (r+R_s)^2-\rho_0 R_s^3\right)}{\rho_c r^2 (r+R_s)^4}.\label{Econd4_model1}
\end{eqnarray}
Analyzing Eq. \eqref{Econd1_model1}, we notice that, for $\omega > -1$, NEC will be violated for small values of the radial coordinate. This violation can be mitigated by increasing the value of $\rho_c$ or decreasing the value of $\rho_0$. As long as $\omega > -1$, the value of $\omega$ will not affect the region where NEC is satisfied. If $\omega < -1$, the regions where NEC was previously satisfied will now be violated, and where it was violated, it will now be satisfied. Through Eq. \eqref{Econd2_model1}, we see that if $\omega > -\frac{1}{3}$, the relation $\rho_e + 3p_e$ will be positive for regions farther from the wormhole, thus the SEC will be violated in more central regions. For $-\frac{1}{3} \geq \omega \geq -\frac{2}{3}$, the SEC will always be violated. Finally, for $\omega < -\frac{2}{3}$, the SEC will be violated in more distant regions and satisfied in regions closer to the wormhole. From Eq. \eqref{Econd3_model1}, we see that for $\omega \leq 1$, the relation $\rho_e - p_e$ will always be positive. This ensures that at least one of the inequalities of DEC will always be satisfied. If $\omega > 1$, the relation will be negative for points farther from the wormhole and positive in more central regions. The region where this inequality will be violated depends on the value of $\omega$; the larger $\omega$ is, the larger the region where the DEC will be violated. The behavior of Eq. \eqref{Econd4_model1} is similar to that of Eq. \eqref{Econd1_model1}. The density $\rho_e$ will be negative for small values of the radial coordinate and positive for points farther from the wormhole. This means that one of the inequalities of the WEC will not be satisfied in more central regions. This violation can be mitigated by increasing $\rho_c$ or decreasing the values of $\rho_0$.

The quantity of exotic matter for the NFW model, calculated via Eq. (\ref{viq}) yields
\begin{eqnarray}
\mathcal{I}_v&=&4 \pi R_s^3 (1 + w) \frac{\rho_0}{\rho_c} \Bigg[ 
    \frac{2}{3} R_s^3 \left( \frac{1}{(r + R_s)^3} - \frac{1}{(r_0 + R_s)^3} \right) \rho_0 + R_s \left( \frac{1}{(r + R_s)} - \frac{1}{(r_0 + R_s)} \right) \rho_c \nonumber \\
    &+& \rho_c \log \left( \frac{r + R_s}{r_0 + R_s} \right) 
\Bigg].
\end{eqnarray}
In the limit $r\gg R_s$, we have $\mathcal{I}_v\approx 4 \pi  \rho_0 R_s^3 (\omega+1) \log \left(\frac{r}{r_0+R_s}\right)$. Then, when $r\to \infty$ the integral diverges logarithmically. However, for $\omega\to-1$ this quantity vanishes. 

\subsection{PI model}
Considering the model \eqref{piDens}, we can calculate the components of the stress-energy tensor, and we obtain that their combinations are given by:

\begin{eqnarray}
&&\rho_e+p_{e}=\frac{\rho_0 R_s^2 (\omega +1) \left(R_s^2 (\rho_c-2 \rho_0)+\rho_c r^2\right)}{\rho_c \left(r^2+R_s^2\right)^2},\label{Econd1_model2} \\
&&\rho_e+3p_e=\frac{\rho_0 R_s^2 \left(\rho_c r^2 (3 \omega +1)+R_s^2 (-2 \rho_0 (3 \omega +2)+\rho_c(3 \omega +1))\right)}{\rho_c \left(r^2+R_s^2\right)^2},\label{Econd2_model2}\\
&& \rho_e-p_{e}=\frac{\rho_0 R_s^2 \rho_c\left((r^2+R_s^2) (1- \omega )+ 2 R_s^2\rho_0 \omega \right)}{\rho_c \left(r^2+R_s^2\right)^2},\label{Econd3_model2}\\
&&\rho_e=\frac{\rho_0 R_s^2 \left(\rho_c r^2+R_s^2 (\rho_c-\rho_0)\right)}{\rho_c \left(r^2+R_s^2\right)^2}.\label{Econd4_model2}
\end{eqnarray}
For $\omega = -1$, the NEC is identically satisfied, since Eq. \eqref{Econd1_model2} would be null in this case. For $\omega > -1$, NEC will always be satisfied if $\rho_c \geq 2\rho_0$. If $\rho_c < 2\rho_0$, NEC will be violated near the center of the wormhole. This violation will be mitigated the smaller $\rho_0$ is or the larger $\rho_c$ is. For $\rho_c > 2 \rho_0 (3 \omega + 2)/(3 \omega + 1)$, the relation $\rho_e + 3p_e$, Eq. \eqref{Econd2_model2}, will always be positive, thus satisfying SEC. If $\omega < 1$, the relation $\rho_e - p_e$, Eq. \eqref{Econd3_model2}, will always be positive, satisfying DEC. For $\omega > 1$, DEC will be satisfied in more internal regions and violated at points farther from the wormhole.
If $\rho_c > \rho_0$, Eq. \eqref{Econd4_model2}, the energy density $\rho_e$ will always be positive, contributing to NEC being always satisfied.

The quantity of exotic matter at $r$ for the PI model is given by
\begin{eqnarray}    
\mathcal{I}_v&=&4 \pi R_s^2 (1 + \omega) \frac{\rho_0}{\rho_c} \left[ r R_s^2 \rho_0\left(\frac{1}{r^2 + R_s^2}-\frac{1}{r_0^2 + R_s^2}\right) + (r-r_0) \rho_c\right.\nonumber\\
&-&\left. R_s \left( \rho_0 + \rho_c \right) \left(\tan^{-1}\left(\frac{r}{R_s}\right) -\tan^{-1}\left(\frac{r_0}{R_s}\right)\right)\right],
\end{eqnarray}
and for $r\ll R_s$, $\mathcal{I}_v\approx 4\pi R_s^2\rho_0(1+\omega) r$, diverging, therefore, linearly with $r$. Once more, the quantifier vanishes for $\omega\to -1$.

\subsection{PF model}
Considering the model \eqref{pfDens}, we can calculate the components of the stress-energy tensor, and we obtain that their combinations are given by:
\begin{eqnarray}
&&\rho_e+p_{e}=\frac{\rho_0 R_s^3 (\omega +1) \left(\rho_c r^3-2 \rho_0
   R_s^3\right)}{\rho_c r^6},\label{Econd1_model3} \\
&&\rho_e+3p_e=\frac{\rho_0 R_s^3 \left(\rho_c r^3 (3 \omega +1)-2 \rho_0 R_s^3 (3 \omega
   +2)\right)}{\rho_c r^6},\label{Econd2_model3}\\
&& \rho_e-p_{e}=\frac{\rho_0 R_s^3 \left(r^3 \rho_c(1-\omega )+2 \rho_0 R_s^3
   \omega \right)}{\rho_c r^6},\label{Econd3_model3}\\
&&\rho_e=\frac{\rho_0 R_s^3 \left(\rho_c r^3-\rho_0 R_s^3\right)}{\rho_c r^6}.\label{Econd4_model3}
\end{eqnarray}
For $\omega > -1$, the combination $\rho_e + p_e$, Eq. \eqref{Econd1_model3}, will present positive values for $r > \sqrt[3]{2\rho_0/\rho_c}  \, R_s$. Therefore, NEC will only be violated in the region $r < \sqrt[3]{2\rho_0/\rho_c}  \, R_s$. For the case where $\omega < -1$, NEC will be violated for $r >  \sqrt[3]{2\rho_0/\rho_c}  \, R_s$. If $\omega = -1$, NEC will be identically satisfied in all regions. For $\omega > -\frac{1}{3}$, the combination $\rho_e + 3p_e$, Eq. \eqref{Econd2_model3}, is negative for small values of $r$. Thus, SEC is violated in this region, and this violation can be mitigated by increasing $\rho_c$ or decreasing $\rho_0$. For $-\frac{1}{3} \geq \omega > -\frac{2}{3}$, SEC will always be violated. For $\omega < -\frac{2}{3}$, SEC will be violated at points farther from the wormhole, with this violation being mitigated by increasing $\rho_0$ or decreasing $\rho_c$. For $\omega \leq 1$, the combination $\rho_e - p_e$, \eqref{Econd3_model3}, will always be positive, so the DEC will always be satisfied. For $\omega > 1$, the DEC will be violated in regions farther from the wormhole. For $r > \sqrt[3]{\rho_0/\rho_c} \, R_s$, the energy density $\rho_e$ is always positive, so WEC will be violated in regions where $r < \sqrt[3]{\rho_0/\rho_c} \, R_s$.

For the PF dark matter model, the quantifier of exotic dark matter is
\begin{equation}
  \mathcal{I}_v=  4 \pi R_s^3 (1 + \omega) \frac{\rho_0}{\rho_c} \left[ \frac{2 R_s^3 \rho_0}{3}\left(\frac{1}{ r^3} -\frac{1}{r_0^3}\right) + \rho_c \log\left(\frac{r}{r_0}\right) \right],
\end{equation}
which also diverges logarithmically for $r\to \infty$ and vanishes for $\omega\to -1$.
\begin{figure}
    \centering
    \includegraphics[width=0.55\linewidth]{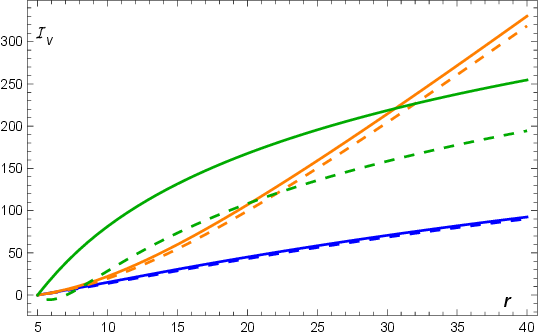}
    \caption{VIQ as a function of the radial coordinate, for the NFW model, in blue and dashed blue, for PI model, in green and dashed green, and for PF model, in orange and dashed orange. The set parameters are $\rho_c=1.0$ (solid curves), $\rho_c=0.1$ (dashed curves), $\rho_0=0.01$, $r_0=5.0$, $\omega=-1/2$, and $R_s=10.0$.
}
       \label{fig_viq}
\end{figure}

In Figure \ref{fig_viq}, we show VIQ as a function of the radial coordinate for the three models, each evaluated at two different values of $\rho_c$. Notice that as this latter decreases, the quantity of exotic matter reduces across all models, with the PF model exhibiting the most significant reduction. Overall, the PI model requires the largest amount of exotic matter, while the NFW model requires the least.

In LQC, a lower critical density \(\rho_c\) introduces quantum gravitational effects at lower densities, reducing the need for exotic matter to sustain non-standard spacetime structures. Additionally, quantum corrections impact spacetime at larger radial distances, stabilizing structures like wormholes without as much exotic matter. Thus, a reduced \(\rho_c\) shifts the balance, allowing quantum geometry to fulfill part of the role typically assigned to exotic dark matter.

\section{Shadow of the LQC wormhole with NFW profile}

We proceed now with the analysis of the wormhole shadow, whose radius is given by  
 \begin{equation}  
    R_{sh} \approx R_o \sin{\alpha_{sh}},  
\end{equation}  
for a distant observer located at \( R_o \) \cite{Alloqulov:2024olb}. The angle \( \alpha_{sh} \) is expressed from:  
\begin{equation}  
\sin{\alpha_{sh}} = \frac{\gamma(r_{ph})}{\gamma(R_o)},  
\end{equation}  
where  
\begin{equation}  
\gamma(r) = \sqrt{-\frac{g^{tt}}{g^{\phi\phi}}}. 
\end{equation}  
 The photon sphere radius, \( r_{ph} \), can be determined by solving:  
\begin{equation}  
    \frac{d\gamma^2(r)}{dr} = 0 \quad \text{at} \quad r = r_{ph}.  
\end{equation}  
Here we will consider $r_{ph}=r_0$, since the calculation will be made in the approximation of a vacuum medium, {\it i.e.}, without a plasma disc around the object, which would affect the photon sphere. For our wormhole solution whose metric coefficient $g_{tt}$ is given by Eq. (\ref{NFWredshift}), where $g_{\phi\phi} = r^2$ at the equatorial plane, the shadow radius becomes:  
\begin{equation}  \label{shadow}
    R_{sh} \approx r_0 \left( \frac{R_o}{r_0} \right)^{\frac{1 + 2w}{1 + w}}
\left( \frac{R_o + R_s}{r_0 + R_s} \right)^{\frac{2(1 + 2w)}{1 + w}}
\left[ \frac{r_0 (r_0 + R_s)^2 \rho_c - 2 R_s^3 \rho_0}{R_o (R_o + R_s)^2 \rho_c - 2 R_s^3 \rho_0} \right].
\end{equation} 
\begin{figure}
    \centering
    \includegraphics[width=0.75\linewidth]{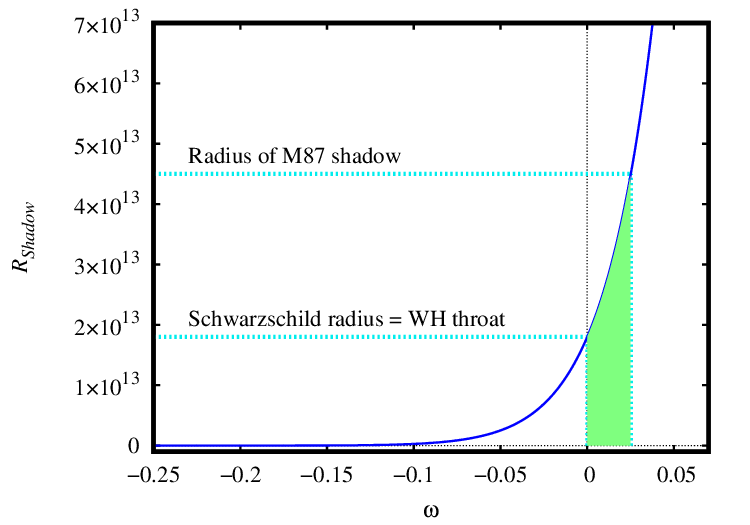}
    \caption{Shadow radius, in meters, as a function of the state parameter $\omega$, considering the following data for the M87 galaxy: $R_o=4.9 \times 10^{23}$ m; $r_0=1.8 \times 10^{13}$ m; $R_s=6.17 \times 10^{20}$ m; $\rho_0=4.4 \times 10^{-21}$ kg/m$^3$; $\rho_c=10^{10}$ kg/m$^3$.}
    \label{figshadow}
\end{figure}
Figure \ref{figshadow} illustrates that the values of the state parameters that yield central shadow sizes consistent with EHT observations fall within the narrow interval $(0.00,0.025)$. Notably, for $\omega\approx 0.025$, the shadow of the wormhole would be indistinguishable from the EHT measurements \cite{EventHorizonTelescope:2019dse}, making it, in principle, impossible to differentiate between a black hole and a wormhole based only on the shadow size. In regions of strong gravitational curvature, such as the vicinity of the supermassive compact object in M87 center, dark matter described by the NFW profile may acquire an effective equation of state with $\omega \neq 0$ due to quantum gravitational effects and the velocity dispersion of dark matter particles. It is worth noting that increasing the critical density $\rho_c$ beyond the value that we have used does not alter our findings and conclusions.

\section{Conclusions}
In this paper, we obtain wormhole solutions in the context of LQC. As a source of matter, we chose three different dark matter profiles that are well-known and studied in the literature. To obtain these solutions, we solved the field equations of LQC and the conservation equation for the energy-momentum tensor, with the appropriate boundary conditions. It was also necessary to choose an equation of state, for which we adopted $p = \omega \rho$. Thus, we have obtained the appropriate expressions for the shape function, $b(r)$, and the redshift function, $\Phi(r)$.

Following, we verified whether the shape function satisfied certain geometric conditions, finding that all models satisfy the established criteria for wormholes. Furthermore, we found that for these spacetimes to be regular, the energy density of dark matter, $\rho_0$, needed to have a maximum limit, with each model having a different limit. Once the criteria for the regularity of the spacetimes were established, we verified through the Kretschmann calculation that there were no singularities in our spacetimes.

Through the embedding diagrams, we were able to visualize the structure of these wormholes. We verified that the smaller the value of $\rho_c$ or the larger the value of $\rho_0$, the faster these wormholes flatten out. The embedding diagrams also allowed us to visualize that there is a minimum radius, which is precisely the throat radius of the wormholes.

We studied the energy conditions for all models. If $\omega = -1$, the null energy condition is identically satisfied for all models. For $\omega \neq -1$, models \eqref{nfwDens} and \eqref{pfDens} always violate the null energy condition in the innermost regions of the wormhole, equations \eqref{Econd1_model1} and \eqref{Econd1_model3}. For the second model, if $\rho_c \geq 2\rho_0$, the null condition will always be satisfied, while for $\rho_c < 2\rho_0$, the null condition will be violated in more central regions, equation \eqref{Econd1_model2}. In models \eqref{nfwDens} and \eqref{pfDens}, the energy densities will always have negative values in central regions, equations \eqref{Econd4_model1} and \eqref{Econd4_model3}. In model \eqref{piDens}, if $\rho_c \geq \rho_0$, the density will be positive at all points, equation \eqref{Econd4_model2}. The second model satisfies the energy conditions with fewer restrictions than the other cases.

We further investigated the amount of exotic dark matter required to stabilize the wormholes by calculating the volume integral quantifier (VIQ), \(\mathcal{I}_v \). Our findings suggest that, across the three analyzed models, a smaller amount of exotic dark matter is generally required as quantum effects intensify (i.e., for lower values of the critical density, \( \rho_c \)), with this effect being particularly notable in the PF model. Overall, the PI model requires the largest amount of exotic matter, while the
NFW model requires the least. However, near the throat, the PF model demands the highest amount of exotic matter, as shown in the plot in Figure \ref{fig_viq}.

Finally, we have applied our result concerning the LQC wormhole sourced by dark matter in the study of the M87 shadow associated with the compact object in its center. Our analysis considered the NFW profile and indicated that, for the state parameter \( \omega \approx 0.025 \), the wormhole's shadow closely aligns with the EHT observations of M87, making it challenging to distinguish between a black hole and a wormhole based solely on shadow imaging. Future research should consider incorporating rotation and surrounding plasma disks to enhance the model's realism, facilitating more precise comparisons with astrophysical data.

In conclusion, we have demonstrated that various dark matter profiles in Loop Quantum Cosmology lead to stable, regular traversable wormhole solutions, with our results highlighting the substantial impact of quantum effects on wormhole structure. This work lays the groundwork for further investigation into dark matter's role in supporting exotic geometries within quantum gravity frameworks.

\section*{Acknowledgments}
\hspace{0.5cm} The authors would like to thank 
Funda\c c\~ao Cearense de Apoio ao Desenvolvimento Cient\'ifico e Tecnol\'ogico (FUNCAP) 
for partial financial support. CRM and GA thanks the Conselho Nacional de Desenvolvimento Cient\'{i}fico e Tecnol\'{o}gico (CNPq), Grants no. 308268/2021-6 and 315568/2021-6.



\bibliography{Ref.bib}

\begin{thebibliography}{35}
\expandafter\ifx\csname natexlab\endcsname\relax\def\natexlab#1{#1}\fi
\expandafter\ifx\csname bibnamefont\endcsname\relax
  \def\bibnamefont#1{#1}\fi
\expandafter\ifx\csname bibfnamefont\endcsname\relax
  \def\bibfnamefont#1{#1}\fi
\expandafter\ifx\csname citenamefont\endcsname\relax
  \def\citenamefont#1{#1}\fi
\expandafter\ifx\csname url\endcsname\relax
  \def\url#1{\texttt{#1}}\fi
\expandafter\ifx\csname urlprefix\endcsname\relax\def\urlprefix{URL }\fi
\providecommand{\bibinfo}[2]{#2}
\providecommand{\eprint}[2][]{\url{#2}}

\bibitem[{\citenamefont{Abbott et~al.}(2016)}]{LIGOScientific:2016aoc}
\bibinfo{author}{\bibfnamefont{B.~P.} \bibnamefont{Abbott}} \bibnamefont{et~al.} (\bibinfo{collaboration}{LIGO Scientific, Virgo}), \bibinfo{journal}{Phys. Rev. Lett.} \textbf{\bibinfo{volume}{116}}, \bibinfo{pages}{061102} (\bibinfo{year}{2016}), \eprint{1602.03837}.

\bibitem[{\citenamefont{Einstein and Rosen}(1935)}]{Einstein:1935tc}
\bibinfo{author}{\bibfnamefont{A.}~\bibnamefont{Einstein}} \bibnamefont{and} \bibinfo{author}{\bibfnamefont{N.}~\bibnamefont{Rosen}}, \bibinfo{journal}{Phys. Rev.} \textbf{\bibinfo{volume}{48}}, \bibinfo{pages}{73} (\bibinfo{year}{1935}).

\bibitem[{\citenamefont{Morris and Thorne}(1988)}]{Morris:1988cz}
\bibinfo{author}{\bibfnamefont{M.~S.} \bibnamefont{Morris}} \bibnamefont{and} \bibinfo{author}{\bibfnamefont{K.~S.} \bibnamefont{Thorne}}, \bibinfo{journal}{Am. J. Phys.} \textbf{\bibinfo{volume}{56}}, \bibinfo{pages}{395} (\bibinfo{year}{1988}).

\bibitem[{\citenamefont{Hawking}(1988)}]{Hawking:1988ae}
\bibinfo{author}{\bibfnamefont{S.~W.} \bibnamefont{Hawking}}, \bibinfo{journal}{Phys. Rev. D} \textbf{\bibinfo{volume}{37}}, \bibinfo{pages}{904} (\bibinfo{year}{1988}).

\bibitem[{\citenamefont{Visser}(1995)}]{Visser:1995cc}
\bibinfo{author}{\bibfnamefont{M.}~\bibnamefont{Visser}}, \emph{\bibinfo{title}{{Lorentzian wormholes: From Einstein to Hawking}}} (\bibinfo{year}{1995}), ISBN \bibinfo{isbn}{978-1-56396-653-8}.

\bibitem[{\citenamefont{Nandi et~al.}(2004)\citenamefont{Nandi, Zhang, and Vijaya~Kumar}}]{Nandi:2004ku}
\bibinfo{author}{\bibfnamefont{K.~K.} \bibnamefont{Nandi}}, \bibinfo{author}{\bibfnamefont{Y.-Z.} \bibnamefont{Zhang}}, \bibnamefont{and} \bibinfo{author}{\bibfnamefont{K.~B.} \bibnamefont{Vijaya~Kumar}}, \bibinfo{journal}{Phys. Rev. D} \textbf{\bibinfo{volume}{70}}, \bibinfo{pages}{127503} (\bibinfo{year}{2004}), \eprint{gr-qc/0407079}.

\bibitem[{\citenamefont{Churilova et~al.}(2021)\citenamefont{Churilova, Konoplya, Stuchlik, and Zhidenko}}]{Churilova:2021tgn}
\bibinfo{author}{\bibfnamefont{M.~S.} \bibnamefont{Churilova}}, \bibinfo{author}{\bibfnamefont{R.~A.} \bibnamefont{Konoplya}}, \bibinfo{author}{\bibfnamefont{Z.}~\bibnamefont{Stuchlik}}, \bibnamefont{and} \bibinfo{author}{\bibfnamefont{A.}~\bibnamefont{Zhidenko}}, \bibinfo{journal}{JCAP} \textbf{\bibinfo{volume}{10}}, \bibinfo{pages}{010} (\bibinfo{year}{2021}), \eprint{2107.05977}.

\bibitem[{\citenamefont{Konoplya and Zhidenko}(2022)}]{Konoplya:2021hsm}
\bibinfo{author}{\bibfnamefont{R.~A.} \bibnamefont{Konoplya}} \bibnamefont{and} \bibinfo{author}{\bibfnamefont{A.}~\bibnamefont{Zhidenko}}, \bibinfo{journal}{Phys. Rev. Lett.} \textbf{\bibinfo{volume}{128}}, \bibinfo{pages}{091104} (\bibinfo{year}{2022}), \eprint{2106.05034}.

\bibitem[{\citenamefont{Ashtekar}(2009)}]{Ashtekar:2008ay}
\bibinfo{author}{\bibfnamefont{A.}~\bibnamefont{Ashtekar}}, \bibinfo{journal}{J. Phys. Conf. Ser.} \textbf{\bibinfo{volume}{189}}, \bibinfo{pages}{012003} (\bibinfo{year}{2009}), \eprint{0812.4703}.

\bibitem[{\citenamefont{Sengupta et~al.}(2023)\citenamefont{Sengupta, Ghosh, and Kalam}}]{Sengupta:2023yof}
\bibinfo{author}{\bibfnamefont{R.}~\bibnamefont{Sengupta}}, \bibinfo{author}{\bibfnamefont{S.}~\bibnamefont{Ghosh}}, \bibnamefont{and} \bibinfo{author}{\bibfnamefont{M.}~\bibnamefont{Kalam}}, \bibinfo{journal}{Eur. Phys. J. C} \textbf{\bibinfo{volume}{83}}, \bibinfo{pages}{830} (\bibinfo{year}{2023}), \eprint{2309.12527}.

\bibitem[{\citenamefont{Muniz et~al.}(2024)\citenamefont{Muniz, Tangphati, Neves, and Cruz}}]{Muniz:2024jzg}
\bibinfo{author}{\bibfnamefont{C.~R.} \bibnamefont{Muniz}}, \bibinfo{author}{\bibfnamefont{T.}~\bibnamefont{Tangphati}}, \bibinfo{author}{\bibfnamefont{R.~M.~P.} \bibnamefont{Neves}}, \bibnamefont{and} \bibinfo{author}{\bibfnamefont{M.~B.} \bibnamefont{Cruz}}, \bibinfo{journal}{Phys. Dark Univ.} \textbf{\bibinfo{volume}{46}}, \bibinfo{pages}{101673} (\bibinfo{year}{2024}), \eprint{2406.08250}.

\bibitem[{\citenamefont{Aghanim et~al.}(2020)}]{Planck:2018vyg}
\bibinfo{author}{\bibfnamefont{N.}~\bibnamefont{Aghanim}} \bibnamefont{et~al.} (\bibinfo{collaboration}{Planck}), \bibinfo{journal}{Astron. Astrophys.} \textbf{\bibinfo{volume}{641}}, \bibinfo{pages}{A6} (\bibinfo{year}{2020}), \bibinfo{note}{[Erratum: Astron.Astrophys. 652, C4 (2021)]}, \eprint{1807.06209}.

\bibitem[{\citenamefont{Zwicky}(1933)}]{Zwicky:1933gu}
\bibinfo{author}{\bibfnamefont{F.}~\bibnamefont{Zwicky}}, \bibinfo{journal}{Helv. Phys. Acta} \textbf{\bibinfo{volume}{6}}, \bibinfo{pages}{110} (\bibinfo{year}{1933}).

\bibitem[{\citenamefont{Rubin and Ford}(1970)}]{Rubin:1970zza}
\bibinfo{author}{\bibfnamefont{V.~C.} \bibnamefont{Rubin}} \bibnamefont{and} \bibinfo{author}{\bibfnamefont{W.~K.} \bibnamefont{Ford}, \bibfnamefont{Jr.}}, \bibinfo{journal}{Astrophys. J.} \textbf{\bibinfo{volume}{159}}, \bibinfo{pages}{379} (\bibinfo{year}{1970}).

\bibitem[{\citenamefont{Persic et~al.}(1996)\citenamefont{Persic, Salucci, and Stel}}]{Persic:1995ru}
\bibinfo{author}{\bibfnamefont{M.}~\bibnamefont{Persic}}, \bibinfo{author}{\bibfnamefont{P.}~\bibnamefont{Salucci}}, \bibnamefont{and} \bibinfo{author}{\bibfnamefont{F.}~\bibnamefont{Stel}}, \bibinfo{journal}{Mon. Not. Roy. Astron. Soc.} \textbf{\bibinfo{volume}{281}}, \bibinfo{pages}{27} (\bibinfo{year}{1996}), \eprint{astro-ph/9506004}.

\bibitem[{\citenamefont{Bertone and Hooper}(2018)}]{Bertone:2016nfn}
\bibinfo{author}{\bibfnamefont{G.}~\bibnamefont{Bertone}} \bibnamefont{and} \bibinfo{author}{\bibfnamefont{D.}~\bibnamefont{Hooper}}, \bibinfo{journal}{Rev. Mod. Phys.} \textbf{\bibinfo{volume}{90}}, \bibinfo{pages}{045002} (\bibinfo{year}{2018}), \eprint{1605.04909}.

\bibitem[{\citenamefont{Randall}(2018)}]{randall2018}
\bibinfo{author}{\bibfnamefont{L.}~\bibnamefont{Randall}}, \bibinfo{journal}{Nature} \textbf{\bibinfo{volume}{557}}, \bibinfo{pages}{2} (\bibinfo{year}{2018}).

\bibitem[{\citenamefont{Arg\"uelles et~al.}(2021)\citenamefont{Arg\"uelles, Diaz, Kheirandish, Olivares-Del-Campo, Safa, and Vincent}}]{Arguelles:2019ouk}
\bibinfo{author}{\bibfnamefont{C.~A.} \bibnamefont{Arg\"uelles}}, \bibinfo{author}{\bibfnamefont{A.}~\bibnamefont{Diaz}}, \bibinfo{author}{\bibfnamefont{A.}~\bibnamefont{Kheirandish}}, \bibinfo{author}{\bibfnamefont{A.}~\bibnamefont{Olivares-Del-Campo}}, \bibinfo{author}{\bibfnamefont{I.}~\bibnamefont{Safa}}, \bibnamefont{and} \bibinfo{author}{\bibfnamefont{A.~C.} \bibnamefont{Vincent}}, \bibinfo{journal}{Rev. Mod. Phys.} \textbf{\bibinfo{volume}{93}}, \bibinfo{pages}{035007} (\bibinfo{year}{2021}), \eprint{1912.09486}.

\bibitem[{\citenamefont{Marsh et~al.}(2024)\citenamefont{Marsh, Ellis, and Mehta}}]{Marsh:2024ury}
\bibinfo{author}{\bibfnamefont{D.~J.~E.} \bibnamefont{Marsh}}, \bibinfo{author}{\bibfnamefont{D.}~\bibnamefont{Ellis}}, \bibnamefont{and} \bibinfo{author}{\bibfnamefont{V.~M.} \bibnamefont{Mehta}}, \emph{\bibinfo{title}{{Dark Matter: Evidence, Theory, and Constraints}}}, Princeton Series in Astrophysics (\bibinfo{publisher}{Princeton University Press}, \bibinfo{year}{2024}), ISBN \bibinfo{isbn}{978-0-691-24971-1, 978-0-691-24952-0}.

\bibitem[{\citenamefont{Tsai et~al.}(2023)\citenamefont{Tsai, Eby, and Safronova}}]{Tsai:2021lly}
\bibinfo{author}{\bibfnamefont{Y.-D.} \bibnamefont{Tsai}}, \bibinfo{author}{\bibfnamefont{J.}~\bibnamefont{Eby}}, \bibnamefont{and} \bibinfo{author}{\bibfnamefont{M.~S.} \bibnamefont{Safronova}}, \bibinfo{journal}{Nature Astron.} \textbf{\bibinfo{volume}{7}}, \bibinfo{pages}{113} (\bibinfo{year}{2023}), \eprint{2112.07674}.

\bibitem[{\citenamefont{Souza et~al.}(2025)\citenamefont{Souza, Muniz, Neves, and Cruz}}]{Souza:2024ltj}
\bibinfo{author}{\bibfnamefont{A.~D.~S.} \bibnamefont{Souza}}, \bibinfo{author}{\bibfnamefont{C.~R.} \bibnamefont{Muniz}}, \bibinfo{author}{\bibfnamefont{R.~M.~P.} \bibnamefont{Neves}}, \bibnamefont{and} \bibinfo{author}{\bibfnamefont{M.~B.} \bibnamefont{Cruz}}, \bibinfo{journal}{Annals Phys.} \textbf{\bibinfo{volume}{472}}, \bibinfo{pages}{169859} (\bibinfo{year}{2025}), \eprint{2403.04992}.

\bibitem[{\citenamefont{Xu et~al.}(2020)\citenamefont{Xu, Tang, Cao, and Zhang}}]{Xu:2020wfm}
\bibinfo{author}{\bibfnamefont{Z.}~\bibnamefont{Xu}}, \bibinfo{author}{\bibfnamefont{M.}~\bibnamefont{Tang}}, \bibinfo{author}{\bibfnamefont{G.}~\bibnamefont{Cao}}, \bibnamefont{and} \bibinfo{author}{\bibfnamefont{S.-N.} \bibnamefont{Zhang}}, \bibinfo{journal}{Eur. Phys. J. C} \textbf{\bibinfo{volume}{80}}, \bibinfo{pages}{70} (\bibinfo{year}{2020}).

\bibitem[{\citenamefont{Muniz and Maluf}(2022)}]{Muniz:2022eex}
\bibinfo{author}{\bibfnamefont{C.~R.} \bibnamefont{Muniz}} \bibnamefont{and} \bibinfo{author}{\bibfnamefont{R.~V.} \bibnamefont{Maluf}}, \bibinfo{journal}{Annals Phys.} \textbf{\bibinfo{volume}{446}}, \bibinfo{pages}{169129} (\bibinfo{year}{2022}).

\bibitem[{\citenamefont{Mustafa et~al.}(2023)\citenamefont{Mustafa, Maurya, and Ray}}]{Mustafa:2023kqt}
\bibinfo{author}{\bibfnamefont{G.}~\bibnamefont{Mustafa}}, \bibinfo{author}{\bibfnamefont{S.~K.} \bibnamefont{Maurya}}, \bibnamefont{and} \bibinfo{author}{\bibfnamefont{S.}~\bibnamefont{Ray}}, \bibinfo{journal}{Fortsch. Phys.} \textbf{\bibinfo{volume}{71}}, \bibinfo{pages}{2200129} (\bibinfo{year}{2023}).

\bibitem[{\citenamefont{Radhakrishnan et~al.}(2024)\citenamefont{Radhakrishnan, Brown, Mutulevich, Davis, Mirfendereski, and Cleaver}}]{Radhakrishnan:2024rnm}
\bibinfo{author}{\bibfnamefont{R.}~\bibnamefont{Radhakrishnan}}, \bibinfo{author}{\bibfnamefont{P.}~\bibnamefont{Brown}}, \bibinfo{author}{\bibfnamefont{J.}~\bibnamefont{Mutulevich}}, \bibinfo{author}{\bibfnamefont{E.}~\bibnamefont{Davis}}, \bibinfo{author}{\bibfnamefont{D.}~\bibnamefont{Mirfendereski}}, \bibnamefont{and} \bibinfo{author}{\bibfnamefont{G.}~\bibnamefont{Cleaver}}, \bibinfo{journal}{Symmetry} \textbf{\bibinfo{volume}{16}}, \bibinfo{pages}{1007} (\bibinfo{year}{2024}), \eprint{2405.05476}.

\bibitem[{\citenamefont{Errehymy et~al.}(2024)\citenamefont{Errehymy, Banerjee, Hansraj, Donmez, Nisar, and Abdel-Aty}}]{Errehymy:2024lhl}
\bibinfo{author}{\bibfnamefont{A.}~\bibnamefont{Errehymy}}, \bibinfo{author}{\bibfnamefont{A.}~\bibnamefont{Banerjee}}, \bibinfo{author}{\bibfnamefont{S.}~\bibnamefont{Hansraj}}, \bibinfo{author}{\bibfnamefont{O.}~\bibnamefont{Donmez}}, \bibinfo{author}{\bibfnamefont{K.~S.} \bibnamefont{Nisar}}, \bibnamefont{and} \bibinfo{author}{\bibfnamefont{A.-H.} \bibnamefont{Abdel-Aty}}, \bibinfo{journal}{Eur. Phys. J. C} \textbf{\bibinfo{volume}{84}}, \bibinfo{pages}{573} (\bibinfo{year}{2024}).

\bibitem[{\citenamefont{Maurya et~al.}(2024)\citenamefont{Maurya, Kumar, Kiroriwal, and Errehymy}}]{Maurya:2024jos}
\bibinfo{author}{\bibfnamefont{S.~K.} \bibnamefont{Maurya}}, \bibinfo{author}{\bibfnamefont{J.}~\bibnamefont{Kumar}}, \bibinfo{author}{\bibfnamefont{S.}~\bibnamefont{Kiroriwal}}, \bibnamefont{and} \bibinfo{author}{\bibfnamefont{A.}~\bibnamefont{Errehymy}}, \bibinfo{journal}{Phys. Dark Univ.} \textbf{\bibinfo{volume}{46}}, \bibinfo{pages}{101564} (\bibinfo{year}{2024}).

\bibitem[{\citenamefont{Hassan and Sahoo}(2024)}]{Hassan:2024xyx}
\bibinfo{author}{\bibfnamefont{Z.}~\bibnamefont{Hassan}} \bibnamefont{and} \bibinfo{author}{\bibfnamefont{P.~K.} \bibnamefont{Sahoo}}, \bibinfo{journal}{Annalen Phys.} \textbf{\bibinfo{volume}{536}}, \bibinfo{pages}{2400114} (\bibinfo{year}{2024}), \eprint{2406.13224}.

\bibitem[{\citenamefont{Navarro et~al.}(1996)\citenamefont{Navarro, Frenk, and White}}]{Navarro:1995iw}
\bibinfo{author}{\bibfnamefont{J.~F.} \bibnamefont{Navarro}}, \bibinfo{author}{\bibfnamefont{C.~S.} \bibnamefont{Frenk}}, \bibnamefont{and} \bibinfo{author}{\bibfnamefont{S.~D.~M.} \bibnamefont{White}}, \bibinfo{journal}{Astrophys. J.} \textbf{\bibinfo{volume}{462}}, \bibinfo{pages}{563} (\bibinfo{year}{1996}), \eprint{astro-ph/9508025}.

\bibitem[{\citenamefont{Begeman et~al.}(1991)\citenamefont{Begeman, Broeils, and Sanders}}]{Begeman:1991iy}
\bibinfo{author}{\bibfnamefont{K.~G.} \bibnamefont{Begeman}}, \bibinfo{author}{\bibfnamefont{A.~H.} \bibnamefont{Broeils}}, \bibnamefont{and} \bibinfo{author}{\bibfnamefont{R.~H.} \bibnamefont{Sanders}}, \bibinfo{journal}{Mon. Not. Roy. Astron. Soc.} \textbf{\bibinfo{volume}{249}}, \bibinfo{pages}{523} (\bibinfo{year}{1991}).

\bibitem[{\citenamefont{Li and Yang}(2012)}]{Li:2012zx}
\bibinfo{author}{\bibfnamefont{M.-H.} \bibnamefont{Li}} \bibnamefont{and} \bibinfo{author}{\bibfnamefont{K.-C.} \bibnamefont{Yang}}, \bibinfo{journal}{Phys. Rev. D} \textbf{\bibinfo{volume}{86}}, \bibinfo{pages}{123015} (\bibinfo{year}{2012}), \eprint{1204.3178}.

\bibitem[{\citenamefont{Ashraf et~al.}(2024)\citenamefont{Ashraf, Javed, Ma, and Waseem}}]{Ashraf:2024bol}
\bibinfo{author}{\bibfnamefont{A.}~\bibnamefont{Ashraf}}, \bibinfo{author}{\bibfnamefont{F.}~\bibnamefont{Javed}}, \bibinfo{author}{\bibfnamefont{W.-X.} \bibnamefont{Ma}}, \bibnamefont{and} \bibinfo{author}{\bibfnamefont{A.}~\bibnamefont{Waseem}}, \bibinfo{journal}{Eur. Phys. J. Plus} \textbf{\bibinfo{volume}{139}}, \bibinfo{pages}{857} (\bibinfo{year}{2024}).

\bibitem[{\citenamefont{Swain et~al.}(2024)\citenamefont{Swain, Sahoo, and Nayak}}]{Swain:2024vnc}
\bibinfo{author}{\bibfnamefont{S.}~\bibnamefont{Swain}}, \bibinfo{author}{\bibfnamefont{G.}~\bibnamefont{Sahoo}}, \bibnamefont{and} \bibinfo{author}{\bibfnamefont{B.}~\bibnamefont{Nayak}}, \bibinfo{journal}{Sci. Rep.} \textbf{\bibinfo{volume}{14}}, \bibinfo{pages}{16928} (\bibinfo{year}{2024}).

\bibitem[{\citenamefont{Alloqulov et~al.}(2024)\citenamefont{Alloqulov, Atamurotov, Abdujabbarov, Ahmedov, and Khamidov}}]{Alloqulov:2024olb}
\bibinfo{author}{\bibfnamefont{M.}~\bibnamefont{Alloqulov}}, \bibinfo{author}{\bibfnamefont{F.}~\bibnamefont{Atamurotov}}, \bibinfo{author}{\bibfnamefont{A.}~\bibnamefont{Abdujabbarov}}, \bibinfo{author}{\bibfnamefont{B.}~\bibnamefont{Ahmedov}}, \bibnamefont{and} \bibinfo{author}{\bibfnamefont{V.}~\bibnamefont{Khamidov}}, \bibinfo{journal}{Chin. Phys. C} \textbf{\bibinfo{volume}{48}}, \bibinfo{pages}{025104} (\bibinfo{year}{2024}).

\bibitem[{\citenamefont{Akiyama et~al.}(2019)}]{EventHorizonTelescope:2019dse}
\bibinfo{author}{\bibfnamefont{K.}~\bibnamefont{Akiyama}} \bibnamefont{et~al.} (\bibinfo{collaboration}{Event Horizon Telescope}), \bibinfo{journal}{Astrophys. J. Lett.} \textbf{\bibinfo{volume}{875}}, \bibinfo{pages}{L1} (\bibinfo{year}{2019}), \eprint{1906.11238}.

\end{thebibliography}
\end{document}